\documentclass[12pt]{article}
\pdfoutput=1
\usepackage[T1]{fontenc}
\usepackage[utf8]{inputenc}
\usepackage{ulem}

\usepackage{color}
\usepackage{bm}
\usepackage{slashed}
\usepackage{varioref}
\usepackage[usenames,dvipsnames]{xcolor}
\usepackage{graphicx}
\usepackage{amsmath}
\usepackage{amssymb}
\usepackage{amsfonts}
\usepackage{amsthm}
\usepackage{array}
\usepackage{yfonts}
\usepackage{float}
\usepackage{comment}
\usepackage{caption}
\usepackage[sort]{cite}
\usepackage{subcaption}
\usepackage{wrapfig}
\usepackage[
      colorlinks=true,
      linkcolor=blue,
      urlcolor=blue,
      filecolor=blue,
      citecolor=red,
      pdfstartview=FitV,
      pdftitle={},
      pdfauthor={},
      pdfsubject={},
      pdfkeywords={},
      bookmarksopen=true
]{hyperref}

\usepackage{epsfig}

\usepackage{hyperref}
\newcommand{\mcN}{{\mycal N}}

\newcommand{\sectionofScri}%
{{ \,\,\,\,\mathring{\!\!\!\!\mcN}}}

\newcommand{\scri}{{\mycal I}}%
\newcommand{\eeal}[1]{\label{#1}\end{eqnarray}}

\DeclareFontFamily{OT1}{rsfs}{}
\DeclareFontShape{OT1}{rsfs}{m}{n}{ <-7> rsfs5 <7-10> rsfs7 <10-> rsfs10}{}
\DeclareMathAlphabet{\mycal}{OT1}{rsfs}{m}{n}

\newcommand{\mmr}[1]{\mnotex{{\bf mm:} {\color{violet} #1}}}

\definecolor{applegreen}{rgb}{0.55, 0.71, 0.0}
\definecolor{armygreen}{rgb}{0.29, 0.33, 0.13}

\definecolor{caribbeangreen}{rgb}{0.0, 0.8, 0.6}

\newcounter{mnotecount}[section]

\renewcommand{\themnotecount}{\thesection.\arabic{mnotecount}}

\newcommand{\mnotex}[1]
{\protect{\stepcounter{mnotecount}}$^{\mbox{\footnotesize
        $
        \bullet$\themnotecount}}$ \marginpar{
    \raggedright\tiny\em
    $\!\!\!\!\!\!\,\bullet$\themnotecount: #1} }

\newcommand{\ptc}[1]{\mnotex{{\bf ptc:} {  #1}}}
\newcommand{\ptcrr}[1]{\mnotex{{\bf ptc:} {  #1}}}

\newcommand{\bel}[1]{\begin{equation}\label{#1}}

\newcommand{\eeq}{\end{equation}}
\newcommand{\ee}{\end{equation}}
\newcommand{\beqa}{\begin{eqnarray}}
\newcommand{\eeqa}{\end{eqnarray}}
\newcommand{\beqan}{\begin{eqnarray*}}
\newcommand{\eeqan}{\end{eqnarray*}}
\newcommand{\ba}{\begin{array}}
\newcommand{\ea}{\end{array}}

\newcommand{\ptcheck}[1]{\ptc{checked on #1}}

\def\beq{\begin{eqnarray}}
\def\eeq{\end{eqnarray}}

\def\a{\alpha}
\def\b{\beta}

\def\be{\begin{equation}}
\def\ee{\end{equation}}
\def\bea{\begin{eqnarray}}
\def\eea{\end{eqnarray}}

\newcommand{\tscomr}[1]{\mnotex{{\bf ts:} {  #1}}}

{\catcode `\@=11 \global\let\AddToReset=\@addtoreset}
\AddToReset{equation}{section}

\renewcommand{\ptcrr}[1]{}

\renewcommand{\mmr}[1]{}
\renewcommand{\tscomr}[1]{}

\renewcommand{\ptcheck}[1]{}

\begin{document}

\title{\bf{Quadrupolar radiation in de Sitter:\\ Displacement memory and Bondi metric}}

\date{}

\maketitle

\vspace{-1.5cm}

\centerline{\large{\bf{Geoffrey Comp\`ere}$^{a}$\footnote{geoffrey.compere@ulb.be},  Sk Jahanur Hoque$^{a,b,c}$\footnote{jahanur.hoque@hyderabad.bits-pilani.ac.in}, Emine \c{S}eyma Kutluk$^{d}$\footnote{kutlukes@itu.edu.tr}}}\vspace{6pt}

\bigskip\medskip
\centerline{\textit{{}$^{a}$ Universit\'{e} Libre de Bruxelles, International Solvay Institutes,}}
\centerline{\textit{CP 231, B-1050 Brussels, Belgium}}

\medskip
\centerline{\textit{{}$^{b}$ Birla Institute of Technology and Science, Pilani, Hyderabad Campus, }}
\centerline{\textit{Jawaharnagar, Hyderabad 500 078, India}}

\medskip
\centerline{\textit{{}$^{c}$ Institute of Theoretical Physics, Faculty of Mathematics and Physics, Charles University, }}
\centerline{\textit{V Holešovičkách 2, 180 00 Prague 8, Czech Republic}}

\medskip
\centerline{\textit{{}$^{d}$ Department of Physics, \.{I}stanbul Technical University, }}
\centerline{\textit{Maslak 34469 \.{I}stanbul, T\"{u}rkiye}}

\vspace{1cm}

\def\c{\gamma}

\begin{abstract}
We obtain the closed form expression for the metric perturbation around de Sitter spacetime generated by a matter source below Hubble scale both in generalized harmonic gauge and in Bondi gauge up to quadrupolar order in the multipolar expansion, including both parities (i.e. both mass and current quadrupoles). We demonstrate that such a source causes a displacement memory effect close to future infinity that originates, in the even-parity sector, from a $\Lambda$-BMS transition between the two non-radiative regions of future infinity. 

\end{abstract}
\newpage
\tableofcontents
\section{Introduction}

How to accurately model gravitational radiation on cosmological scales? The standard model of cosmology is based upon Einstein's equations coupled to matter. Under the assumption of homogeneity and isotropy, Einstein's equations reduce to Friedman-Robertson-Lema\^itre-Walker (FRLW) universes. Such a FRLW universe adequately describes the history of our universe which asymptotes at early and late times to de Sitter universe due, respectively, to inflation \cite{PhysRevD.23.347} and to dark energy \cite{1998AJ....116.1009R,1999ApJ...517..565P}. Primordial fluctuations in the early universe led after inflation to a nearly Gaussian spectrum of fluctuations, which governed the formation of the large scale structure of matter of our universe.
The astrophysical gravitational wave background \cite{PhysRevLett.118.151105} was recently first detected to $3.5\sigma$ confidence level by the NANOGrav Collaboration \cite{2023ApJ...951L...8A}. One potential target of third generation detectors is the observation of the primordial gravitational wave background, which requires a high resolution of individual astrophysical events in order to subtract the astrophysical gravitational wave background. Over their last three runs, the LVK Collaboration has detected binary black hole merger events that have a cosmological redshift up to order one \cite{theligoscientificcollaboration2021gwtc3}. Third generation detectors will push the redshift of observed events  up to 10 or higher \cite{Sathyaprakash:2019nnu,vanSon:2021zpk,Branchesi:2023mws}. A very large number of astrophysical events will be detected simultaneously and precision models will become  necessary to resolve them. On cosmological scales, gravitational wave propagation is influenced by the background curvature of our FRLW universe, which leads to new tail and memory effects 
\cite{Ford:1977dj,Waylen:1978dx,Braginsky:1985vlg,Caldwell:1993xw,Iliopoulos:1998wq,deVega:1998ia,Akhmedov:2010ah,Chu:2016qxp,Tolish:2016ggo,Jokela:2022rhk}. 

The modern description of gravitational radiation produced by localized sources such as binary mergers in asymptotically flat spacetimes relies on the multipolar post-Newtonian/ post-Minkowskian formalism \cite{1986RSPTA.320..379B,Blanchet:1986dk}. This article aims at building foundations for what could be called the ``multipolar post-de Sitter formalism'' that accounts for gravitational wave generation and propagation around de Sitter background in a multipolar expansion. Important results along these lines have been already derived in the literature:  
the linearized field equations were decoupled using a generalized harmonic gauge condition \cite{deVega:1998ia}, the even-parity-quadrupolar formula for radiation was derived \cite{Ashtekar:2014zfa,Ashtekar:2015lxa, Date:2016uzr, Dobkowski-Rylko:2022dva} and an even-parity-quadrupolar displacement memory effect close to future infinity $\scri^+$ was found \cite{Chu:2015yua,Chu:2016qxp}\footnote{Yet, while reproducing these results, we identified some flaws in each of these treatments, as will be described and corrected in the main text.}. Our first objective is to derive a consistent solution for the metric perturbation in terms of moments of the matter stress-energy tensor up to quadrupolar order including both even and odd sectors. The quadrupolar truncation captures the numerically dominant terms and it also allows us to write all relevant dynamical variables explicitly. Going beyond the quadrupolar order is beyond the scope of this paper. Given the mapping between harmonic coordinates and Bondi coordinates achieved in linearized theory around Minkowski spacetime \cite{Blanchet:2020ngx} (see also Section 4.2 of \cite{Chrusciel:2021ttc}), our second objective is to similarly map the quadrupolar linear solution in generalized harmonic gauge to Bondi gauge and derive the metric in closed form\footnote{Such a program was announced but not achieved in \cite{He:2017dzb}.}.

Boundary conditions for asymptotically de Sitter spacetimes that are compatible with radiation reaching $\scri^+$  were found \cite{Compere:2019bua,Compere:2020lrt}. Such boundary conditions admit as asymptotic symmetry group the $\Lambda$-BMS groupoid, which reduces in the flat limit to the generalized BMS group \cite{Barnich:2010eb,Campiglia:2015yka}. In flat spacetime, the displacement memory effect   \cite{Zeldovich:1974aa,1977ApJ...216..610T,1978ApJ...223.1037E,Braginsky:1985vlg,Blanchet1990,Christodoulou:1991cr,Wiseman:1991ss,Th92,Blanchet:1992br} can be understood from a transition between two Minkowski vacua related by a BMS supertranslation \cite{Strominger:2014pwa}. Our third objective is to investigate whether, at quadrupolar order, the displacement memory effect at $\scri^+$  can be understood as a particular $\Lambda$-BMS transition.  We will prove at quadrupolar order that $\Lambda$-BMS transitions indeed occur in the even sector but not in the odd sector. This parallels the flat case, where odd-parity displacement memory, which has been shown to occur in linear gravity coupled to specific matter obeying the dominant energy condition, does not correspond to a BMS transition \cite{Satishchandran:2019pyc}.

The rest of the paper is organized as follows. In Section \ref{sec:lin}, we summarize the status of linearized gravity on a de Sitter background, write the field equations and solve them consistently to quadrupolar order. In Section \ref{sec:Bondi} we map these perturbations to Bondi gauge and obtain the closed-form metric. In Section \ref{sec:memory}, we describe the linear displacement memory effect close to $\scri^+$ both in generalized harmonic gauge and in Bondi gauge. We conclude in Section \ref{sec:ccl}.

\section{Linearised gravity on de Sitter in generalized harmonic gauge}
\label{sec:lin}

\begin{figure}[!hbt]
    \centering
    ~~~~~
        \includegraphics[width=0.5\textwidth]
        {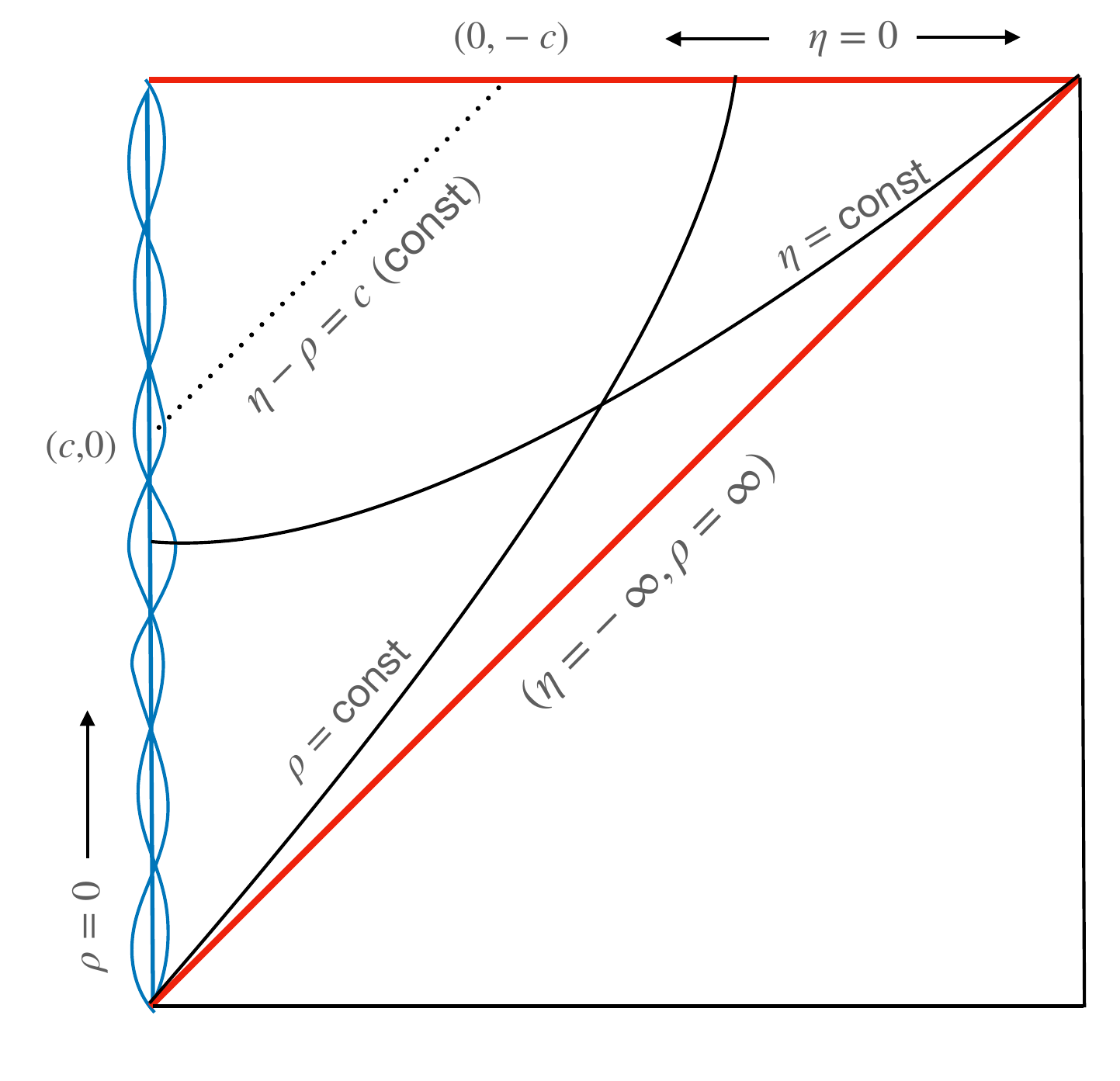}
    \qquad\qquad 
    \caption{Penrose diagram of de Sitter spacetime in conformally flat $(\eta,\rho)$ coordinates, ignoring the angular coordinates. Blue wiggly lines denote sources that are emitting gravitational radiation. Such coordinates  cover the Poincar\'e patch.}\label{PoincarePatch}
\end{figure}

We consider linearised gravity over the de Sitter background. We exclusively work in the future Poincar\'e patch of the de Sitter spacetime, see Figure~\ref{PoincarePatch}. 

 The background de Sitter metric in the Poincar\'e patch is written in conformally flat coordinates $x^\mu=(x^0 = \eta , x^i)$ as 
\be
ds^2 = \bar g_{\a \b} dx^\a dx^\b = a^{2} ( - d\eta^2 + d\vec x^2),  \label{background}
\ee
\be
a=-(H\eta)^{-1}, \quad \mbox{with} \quad H=\sqrt{\Lambda/3},
\ee
where $\Lambda$ is the positive cosmological constant. 
Linearised perturbations over the background \eqref{background} are written as
\be
g_{\a \b} = \bar g_{\a \b} +  h_{\a \b}.
\ee 
We will denote the radial coordinate as $\rho=\sqrt{x^i x^i}$, while we will keep the notation $r$ for the Bondi radial coordinate later on. Coordinates $x_i$, with $i = 1,2,3$, range in $(-\infty, \infty)$, whereas coordinate $
\eta$ takes values in the range $(-\infty, 0)$, with $\eta = 0$ at the future infinity $\scri^+$. The scale factor is positive, $a(\eta)>0$. Future 
infinity $\scri^+$ is a spacelike surface defined as the limit $\eta \rightarrow 0$ at fixed $x^i$. Conformal time $\eta$ is only defined when $\Lambda>0$. In order to take the $\Lambda \to 0$ limit, we will switch to proper time $t$, which is related to conformal time via $\eta = -H^{-1} e^{-Ht}$. 
The Poincar\'e patch of $\scri^+$ covers a $\mathbb R^3$ manifold spanned by $x^i$, which can be understood as the complete $S^3$ manifold $\scri^+$ minus a point, corresponding to the upper right corner of Figure \ref{PoincarePatch}.

The Christoffel symbols of the background metric are 
\be
\overline{\Gamma}^\a_{\b \c} = - \frac{1}{\eta} \left( \delta^0_\b \delta^\a_\c + \delta^0_\c \delta^\a_\b + \delta^\a_0 \eta_{\b \c}\right),
\ee
where $\eta_{\mu\nu}dx^\mu dx^\nu$ is the Minkowski metric.
Using this, useful expressions for the d'Alembertian and for various other derivative operators can be written,  see e.g. \cite{deVega:1998ia, Date:2015kma}.  
In terms of the trace reversed combination 
$\hat{h}_{\alpha\beta}:=h_{\alpha\beta}-\frac{1}{2}\bar{g}
_{\alpha\beta} \ (\bar{g}^{\mu\nu}h_{\mu\nu})$, the linearised Einstein equations take the form,
\begin{equation} 
\frac{1}{2}\left[ - \overline{\Box} \hat{h}_{\mu\nu} + \left\{
\overline{\nabla}_{\mu}B_{\nu} + \overline{\nabla}_{\nu}B_{\mu} -
\bar{g}_{\mu\nu}(\overline{\nabla}^{\alpha}B_{\alpha})\right\}\right] +
\frac{\Lambda}{3}\left[\hat{h}_{\mu\nu} -
\hat{h}\bar{g}_{\mu\nu}\right] ~ = ~ 8\pi G T_{\mu\nu} ,\label{LinEqn}
\end{equation}
where $B_{\mu} := \overline{\nabla}_{\alpha}\hat{h}^{\alpha}_{~\mu}$ and   $\overline{\nabla}_{\alpha}$ is the metric compatible covariant derivative with respect to the background metric $\bar g_{\a \b}$.

As is well known in the literature \cite{deVega:1998ia, Date:2015kma, Ashtekar:2015lxa}, these equations simplify when  written in terms of rescaled variables and in generalized harmonic gauge. We define
\be
\chi_{\mu
\nu}:=a^{-2}\hat{h}_{\mu\nu}. 
\ee 
Using the ``generalized harmonic gauge condition'' \cite{deVega:1998ia},
\be
\partial^{\alpha}\chi_{\alpha\mu} + \frac{1}{\eta}\left(2
\chi_{0\mu} + \delta_{\mu}^0 \chi_{\alpha}^{~\alpha} \right) = 0,
\label{ChiGauge}
\ee
Eq. \eqref{LinEqn} becomes,
\be
-~16 \pi G T_{\mu\nu} = \Box \chi_{\mu\nu} +
\frac{2}{\eta}\partial_0\chi_{\mu\nu} -
\frac{2}{\eta^2}\left(\delta_{\mu}^0\delta_{\nu}^0\chi_{\alpha}^{~\alpha}
+ \delta_{\mu}^0\chi_{0\nu} + \delta_{\nu}^0\chi_{0\mu} \right)
\label{LinEqnChi},
\ee
where $\Box$ is simply the d'Alembertian with respect to Minkowski metric in Cartesian coordinates, $\Box = - \partial_\eta^2 + \partial_i^2$.

In terms of variables 
$
\hat{\chi} :=\chi_{00}+\chi_{ii}, ~\chi_{0i},~ \chi_{ij}
$, 
Eq. (\ref{LinEqnChi}) decomposes into three decoupled equations
\bea \label{11XII.01}
\Box \bigg(\frac{\hat{\chi}}{\eta}\bigg) &=& - \frac{16 \pi G \hat{T}}{\eta}, \label{decoupled1}\\ \label{11XII.02}
\Box \bigg(\frac{{\chi_{0i}}}{\eta}\bigg)&=&- \frac{16 \pi G {T}_{0i}}{\eta}, \label{decoupled2} \\ \label{11XII.03}
\left(\Box + \frac{2}{\eta^2}\right) \left(\frac{\chi_{ij}}{\eta} \right)&=&-\frac{16 \pi G}{\eta} T_{ij},
\label{decoupled3}
\eea
where $\hat{T}:=T_{00}+T_{~i}^{i}$. After solving for $\hat\chi$, $\chi_{0i}$, $\chi_{ij}$ we can reconstruct the metric perturbation from 
\begin{equation}\label{map}
h_{00}=\frac{a^2}{2}\hat \chi, \qquad h_{0i}=a^2 \chi_{0i},\qquad h_{ij}=a^2 \chi_{ij}+a^2\delta_{ij}(\frac{1}{2}\hat \chi - \chi_{ii}). 
\end{equation}

We note that the background metric in Poincar\'e coordinates obeys $\bar g_{\mu\nu}=a^2 \eta_{\mu\nu}$. Under a linearised diffeomorphism  $\xi^{\mu}$, $\chi_{\mu\nu}$ transforms as, 
\be
\delta\chi_{\mu\nu} = (\partial_{\mu}\underline{\xi}_{\nu} +
\partial_{\nu}\underline{\xi}_{\mu} -
\eta_{\mu\nu}\eta^{\alpha\beta}\partial_{\beta}\underline{\xi}_{\alpha}) -
\frac{2}{\eta}\eta_{\mu\nu}\underline{\xi}_0,\label{deltachi1}
\ee
where
\be
\underline{\xi}_{\mu}
:= a^{-2} \xi_{\mu} = \eta_{\mu\nu}\xi^{\nu}.  \label{ChiGaugeTrans}
\ee
 A small calculation then shows that the gauge condition (\ref{ChiGauge}) is preserved under transformations
generated by  vector fields $\xi^{\mu}$ --- the residual gauge transformations --- satisfying
\begin{equation}
\Box\underline{\xi}_{\mu} +
\frac{2}{\eta}\partial_\eta\underline{\xi}_{\mu} -
\frac{2}{\eta^2}\delta_{\mu}^0\underline{\xi}_0= 0,
\label{ResidualChiGauge}
\end{equation}
or, equivalently,\footnote{Note that Eq. \eqref{ResidualChiGauge} matches with Eq. (4.22) of \cite{deVega:1998ia} but Eq. \eqref{Res-Gauge-1} does not match with their Eq. (4.24).}  
\begin{equation}
    \square \left( \frac{\xi^0}{\eta}\right)=0,\qquad \left( \square +\frac{2}{\eta^2} \right) \left( \frac{\xi^i}{\eta}\right)=0. \label{Res-Gauge-1}
\end{equation}
Under these residual gauge transformations
equation \eqref{LinEqnChi} 
is also invariant. 

\subsection{Residual gauge modes and symmetries}
\label{sec:gauge}

In the absence of matter, we can further reduce the gauge freedom as follows. Under a gauge transformation one has 
\begin{eqnarray}
\frac{\delta_\xi \hat \chi}{\eta} &=& -4 \partial_\eta \left( \frac{\xi^0}{\eta} \right), \label{res-gauge-1-}\\
\frac{\delta_\xi \chi_{0i}}{\eta}&=&-\frac{\partial_i \xi^0}{\eta}+ \frac{\partial_\eta \xi^i}{\eta}. \label{res-gauge-2-}
\end{eqnarray}
Applying the D'Alembertian on Eq. \eqref{res-gauge-1-} the left-hand side is zero in the absence of sources after using Eq. \eqref{11XII.01} with $\hat T=0$ and the right-hand side is also zero as a consequence of Eq. \eqref{Res-Gauge-1}. It implies that the harmonic function $\hat\chi/\eta$ can be set to zero by a residual gauge transformation $\xi^0$. The remaining gauge transformation after this gauge fixing is given by $\xi^0=\eta \, \Theta(x^i)$ where $\Delta \Theta \equiv \partial_i \partial_i \Theta =0$. 

Similarly, applying the D'Alembertian on Eq. \eqref{res-gauge-2-} the left-hand side is zero in the absence of sources after using Eq. \eqref{11XII.02} with $T_{0i}=0$ and the right-hand side reduces to 
$\Box \left( \frac{\partial_\eta \xi^i}{\eta}\right)=0$.  One can show that
\begin{equation}
    \Box \left( \frac{\partial_\eta \xi^i}{\eta}\right) = \left( \partial_\eta + \frac{1}{\eta} \right) \left( \Box + \frac{2}{\eta^2} \right) \left( \frac{\xi^i}{\eta}\right) \ .
\end{equation}
Therefore, after using Eq. \eqref{Res-Gauge-1} the right-hand side is also zero. This implies that any $\chi_{0i}$ that solves the homogeneous equations of motion can be set to zero by a gauge transformation $\xi^i$ that satisfies \eqref{Res-Gauge-1}. The remaining gauge transformations are $\xi^i =\Xi^i(x)+\frac{\eta^2}{2}\partial_i \Theta$ with $\Delta \Xi^i=-\partial_i\Theta$. The gauge condition \eqref{ChiGauge} together with $\hat{\chi}=0$ and $\chi_{0i}=0$ implies $\partial_i \chi_{ij}=0$ and $\partial_\eta \chi_{ii}=0$.

Finally, using Eq. \eqref{deltachi1}, one has 
\begin{equation}
    \delta_\xi \chi_{ii}=-\partial_i \underline{\xi}_i+3\partial_0 \underline{\xi}_0-\frac{6}{\eta}\underline{\xi}_0 = - \partial_i \Xi^i + 3 \Theta \ .
\end{equation}
The right hand side is harmonic. Since the homogeneous equation of motion for $\chi_{ii}$ together with $\partial_\eta \chi_{ii}=0$ also enforces $\chi_{ii}$ to be harmonic, any $\chi_{ii}$ that solves the homogeneous equation of motion can be gauge transformed to zero. 

The only remaining degrees of freedom are $\chi^{TT}_{ij}$ the transverse-traceless part of $\chi_{ij}$ obeying \begin{equation}\label{eqgenchiij}
\left(\Box + \frac{2}{\eta^2}\right) \left(\frac{\chi^{TT}_{ij}}{\eta} \right)=0. 
\end{equation}
The final residual gauge freedom 
\begin{eqnarray}
\xi^0 &=& \frac{\eta}{3}\partial_i\Xi^i,\\
\xi^i &=& \Xi^i (\vec{x})+\frac{\eta^2}{6}\partial_i\partial_j\Xi^j    ,
\end{eqnarray}
is parametrized by functions $\Xi^i(\vec{x})$ with $\Delta \Xi^i =-\frac{1}{3}\partial_i\partial_j\Xi^j$, implying $\Delta \partial_i \Xi^i=0$. They transform $\chi^{TT}_{ij}$  as 
\begin{equation} \label{22XI22.01}
\delta_\xi \chi^{TT}_{ij}=\partial_i \Xi^j+\partial_j \Xi^i - \frac{2}{3}\delta_{ij}\partial_k \Xi^k +\frac{\eta^2}{3}\partial_i \partial_j \partial_k \Xi^k.     
\end{equation}
The general solution to \eqref{eqgenchiij} is the sum of a propagating solution (the two polarization modes of gravitational waves) and the residual gauge solution \eqref{22XI22.01}. 

The solutions to $\delta_\xi \chi^{TT}_{ij}=0$ are the 10 Killing symmetries of the de Sitter background that consist of 3 spatial translations, 3 rotations, the dilation and 3 special conformal transformations.

The other solutions  $\Xi^i(\vec{x})$ of $\Delta \Xi^i =-\frac{1}{3}\partial_i\partial_j\Xi^j$ are additional symmetries that are associated to consistency relations \cite{Hinterbichler:2013dpa,Hui:2018cag} (see also \cite{Mirbabayi:2016xvc,Hamada:2018vrw}). The general regular solution is 
\begin{equation}\label{solgensym}
\Xi^i= \sum_{\ell \geq 0} \frac{1}{\ell !} b_{i L}x^L\, ,
\end{equation}
where $b_{i L}$ are constants that are totally symmetric in their last $\ell$ indices and that obey $b_{ijj L-2}=-\frac{1}{3}b_{jijL-2}$ for $\ell \geq 2$\footnote{The 3 $\ell=0$ solutions provide the spatial translations; The 9 $\ell=1$ solutions provide the dilation, the 3 rotations and 5 constant shear shifts $\delta_\xi \chi^{TT}_{ij}=b_{\langle ij \rangle}$. The 15 $\ell=2$ solutions are the 3 special conformal transformations and 12 shear shifts linear in $x^i$, $\delta_\xi \chi^{TT}_{ij}=b_{\langle ij \rangle k} x^k$. All  $\ell \geq 3$ solutions are non-Killing symmetries with $\delta \chi^{TT}_{ij}$ a sum of terms $\sim r^{\ell-1}$ and $\sim \eta^2 r^{\ell-3}$.}. This implies that there is no regular solution to $\Xi^i$ that vanish at infinity\footnote{Contrary to \cite{Ashtekar:2015lxa,Tolish:2016ggo} we do not perform a tensorial decomposition into representations of $SO(3)$, which is valid upon imposing vanishing boundary conditions at infinity. Here, $\chi^{TT}_{ij}$ is not gauge invariant precisely because the residual gauge transformations diverge as $r \rightarrow \infty$.} \footnote{There also exists solutions singular at $r=0$ but the associated Noether charges defined at $r=\infty$ vanish \cite{Compere:2017wrj} and therefore these residual gauge transformations are pure gauge.}. In single field inflation, these residual symmetries become non-linearly realized on the curvature perturbation when considering the spontaneously symmetry breaking of time translations by an inflaton field and they imply the consistency relations of curvature perturbations \cite{Assassi:2012zq}.

\subsection{Moments of the stress-energy tensor}

In order to solve for the inhomogeneous solution we first define and derive properties of the moments of the stress-energy tensor. Integrals of local densities are best defined using tensors expressed in an orthonormal frame tetrad. The frame components are coordinate scalars but transform under the Lorentz group acting on the frame. For the homogeneous metric \eqref{background}, we define the comoving coordinates $\bar x_\alpha=a(\eta)x_\alpha$, the comoving tetrad frame $f^{\alpha}_{\bar \alpha} = a(\eta)^{-1} \delta_{\bar \alpha}^{\alpha}$ defined such as $\eta_{\bar \alpha \bar \beta}= f^{\alpha}_{\bar \alpha}f^{\beta}_{\bar \beta}\bar g_{\alpha\beta}$ and its inverse $f_{ \alpha}^{\bar \alpha} = a(\eta) \delta_{ \alpha}^{\bar \alpha}$. We define the moments of the energy density $Q_L^{(\rho)}$, of the momentum density $P_{i \vert L}$ and of the stress density $S_{ij \vert L}$ as  
\begin{align} 
Q_L^{(\rho)}(\eta)&:=\int d^3 \bar x T_{\bar 0 \bar 0}\bar x_L = \int d^3x a^{\ell + 1}(\eta) T_{00} x_L,\label{25IV23.02} \\ 
P_{i \vert L}(\eta)&:=\int d^3 \bar x T_{\bar 0 \bar i}\bar x_L = \int d^3x a^{\ell + 1}(\eta) T_{0i} x_L, \label{25IV23.02bis}\\
S_{ij \vert L}(\eta) &:=\int d^3 \bar x T_{\bar i \bar j} \bar x_L = \int d^3x a^{\ell + 1}(\eta)  T_{ij}  x_L.\label{27IV23.01}
\end{align} 
We denote $L=i_{1}i_{2}\cdots i_{\ell}$ as a multi-index made of $\ell$ spatial indices, and $x_{L}=x_{i_{1}}\cdots x_{i_{\ell}}$. Such set of moments completely characterize the stress-energy tensor.  For convenience, we also introduce the moments of the pressure density $Q_L^{(p)}$ as
\begin{align} 
\label{25IV23.03}
Q_L^{(p)}(\eta) &=\int d^3 \bar x \eta^{\bar i \bar j}T_{\bar i \bar j} \bar x_L = \int d^3x a^{\ell + 1}(\eta) \delta_{ij}T_{ij} x_L=S_{ii \vert L} ,
\end{align}
and we further define 
\begin{align} 
\label{25IV23.03bis}
Q_L^{(\rho+p)}(\eta) :=Q_L^{(\rho)}(\eta)+Q_L^{(p)}(\eta). 
\end{align}
This extends earlier constructions \cite{Date:2015kma, Ashtekar:2015lxa}. 

The conservation of the stress-energy tensor on the homogeneous spacetime \eqref{background} is equivalent to the flux-balance laws 
\begin{align}
\partial_\eta T_{00}-\partial_i T_{0i}+\frac{a'(\eta)}{a(\eta)}(T_{00}+\delta_{ij}T_{ij}) &=0, \\
\partial_\eta T_{0i}-\partial_j T_{ij}+2 \frac{a'(\eta)}{a(\eta)} T_{0i}&=0. 
\end{align}
The conservation of stress-energy tensor on de Sitter background in conformal coordinates (with $a=-(H\eta)^{-1}$) becomes
\bea 
\partial_\eta T_{00}-\partial_i T_{0i}-\frac{1}{\eta}(T_{00}+\delta_{ij}T_{ij}) &=0, \\
\partial_\eta T_{0i}-\partial_j T_{ij}- \frac{2}{\eta} T_{0i}&=0.
\eea
It is convenient to convert $\partial_{\eta}$ to $\partial_{t}$ using the definition $\eta =- H^{-1} e^{-Ht}$ while remaining in the cosmological coordinates $(\eta,\vec{x})$.  This leads to $\partial_{\eta}=e^{Ht} \partial_{t}:= a(t) \partial_{t}$ and 
\bea   \label{28II23.01}
 a \partial_t T_{00}(\eta,x)-\partial_i T_{0i}(\eta,x)+Ha(T_{00}(\eta,x)+\delta_{ij}T_{ij}(\eta,x)) &=0, \\ \label{28II23.02}
a\partial_t T_{0i}(\eta,x)-\partial_j T_{ij}(\eta,x)+2Ha T_{0i}(\eta,x)&=0.
 \eea
Note that the $0$ index means the $\eta$ component. Multiplying these equations by $a^\ell(\eta)x_L$, integrating over $\vec{x}$ and using the definition of moments of the stress-energy tensor we have
\begin{align}
    \partial_t Q_L^{(\rho)} &= H(\ell Q_L^{(\rho)}-Q_L^{(p)})- \ell  P_{(i_1|i_2 \cdots i_\ell)} \ ,\label{eq37}\\ 
    \partial_t P_{i|L} &= (\ell-1) H P_{i|L} - \ell  S_{i(i_1|i_2 \cdots i_\ell)},\label{eq38}
\end{align}
where $S_{i\emptyset}:=0$ by definition.

For some purposes, it is useful to have handy the first few lowest orders which are given by 
\begin{subequations}
\begin{align}
\partial_{t} Q^{(\rho)}&=-H Q^{(p)},\\
\partial_{t} Q_{i}^{(\rho)}&=-P_{i}+HQ_{i}^{(\rho)}-H Q_{i}^{(p)},\\
\partial_{t} Q^{(\rho)}_{ij}&= -(P_{i|j}+P_{j|i})+2H Q^{(\rho)}_{ij}-H Q^{(p)}_{ij}, \label{Pijd}
\\ \partial_{t} Q^{(\rho)}_{ijk}&= -(P_{i|jk}+P_{j|ki}+P_{k|ij})+3H Q^{(\rho)}_{ijk}-H Q^{(p)}_{ijk},
\end{align}
\end{subequations}
and 
\begin{subequations}
\begin{align}
\partial_{t} P_{i}&=-HP_{i}, \\ \label{27IV23.02first}
\partial_{t}P_{i|j}&=-S_{ij},\\ \label{27IV23.02sec}
\partial_{t} P_{i|jk}&= H P_{i|jk} -S_{ij\vert k}-S_{ik \vert j} ,\\
\partial_{t} P_{i|jkl}&= 2H P_{i|jkl} -S_{ij \vert kl}-S_{ik \vert lj}-S_{il \vert jk} .
\end{align}
\end{subequations}
We define the angular momentum (or odd parity dipole moment) as 
\begin{align}
J_i = \epsilon_{ijk}P_{j \vert k},
\end{align}
which is conserved, $\partial_ t J_i =0$, in the linear theory. 

These expressions can also be written in a coordinate invariant way by substituting $\partial_{t}$ in terms of the Lie derivative of time translation Killing vector field of de Sitter background
\be
T^{\mu} \partial_\mu = - H ( \eta \partial_\eta + \rho \partial_\rho). \label{time_translation}
\ee
Using the relations $\mathcal{L}_{T} Q^{(\rho)}_{L}= \partial_{t} Q^{(\rho)}_{L}- \ell H Q^{(\rho)}_{L}$ and  $\mathcal{L}_{T} P_{i|L}=\partial_t P_{i|L} - (\ell+1) H P_{i|L}$ we have 
\begin{align}
    \mathcal{L}_{T} Q_L^{(\rho)} &= -HQ_L^{(p)}- \ell  P_{(i_1|i_2 \cdots i_\ell)} \ ,\\ 
   \mathcal{L}_{T} P_{i|L} &= -2 H P_{i \vert L} - \ell  S_{i(i_1|i_2 \cdots i_\ell)}.
\end{align}

Using Eqs \eqref{27IV23.02first} and \eqref{Pijd} one obtains
\begin{align} 
S_{ij}(\eta) & =\int d^{3}x' a(\eta)T_{ij} (\eta, x') = \frac{1}{2}\partial_t \left( \partial_t Q_{ij}^{(\rho)} - 2H Q_{ij}^{(\rho)}+H Q_{ij}^{(p)}\right)\label{Sijpartial} \\
& =\frac{1}{2}(\mathcal{L}_{T}^2 Q_{ij}^{(\rho)}+2H\mathcal{L}_{T} Q_{ij}^{(\rho)}+H \mathcal{L}_{T} Q_{ij}^{(p)}+2H^{2} Q_{ij}^{(p)}).\label{Sij}
\end{align}
This reproduces the known formula \cite{Ashtekar:2015lxa, Date:2015kma}. Note that the flat limit towards Minkowski spacetime in standard coordinates $(t,\rho,\theta,\phi)$ is well-defined once one makes the substitutions 
\begin{align}
\mathcal L_T \mapsto \partial_t \qquad 
a \mapsto 1 ,\qquad
H \mapsto 0, \qquad 
 -(H \eta)^{-1} = e^{Ht} \mapsto  1.
\end{align}
All the dependence in $Q^{(p)}_{ij}$ disappears and we recover the standard expression $S_{ij}=\frac{1}{2} \ddot Q_{ij}^{(\rho)}$. 

Solving Eq. \eqref{eq38} with $\ell \mapsto \ell+1$ in terms of $S_{i(j \vert L)}$ we obtain
\begin{align}
S_{i (j \vert L )} &= -\frac{1}{\ell +1}(\partial_t - \ell H)P_{i \vert j L}.\label{eqn37}
\end{align}
The totally symmetric part $S_{(ij \vert L)}$ is a function of $P_{(i\vert j L)}$ which can be substituted using Eq. \eqref{eq37} with $\ell \mapsto \ell + 2$. The two conservation equations \eqref{eq37}-\eqref{eq38} are therefore equivalent to Eq. \eqref{eqn37} and 
\begin{align}
S_{(ij \vert L)} &= \frac{1}{(\ell+1)(\ell+2)}(\partial_t - \ell  H)\left( (\partial_t - (\ell+2)H)Q^{(\rho)}_{ij L}+H Q^{(p)}_{ij L}\right).
\end{align}
Contracting this identity with $\delta_{ij}$ we obtain directly
\begin{equation}
    2 Q_L^{(p)}=(\partial_t -\ell H) (\partial_t Q_{kkL}^{(\rho)}+ H (Q_{kkL}^{(p)}-(\ell+2)Q_{kkL}^{(\rho)})) - \ell (\ell-1) S_{(i_1i_2|...i_\ell)kk} -4\ell S_{k(i_{1}|\cdots i_{\ell})k} .\label{CONSCOND}
\end{equation}
In particular, for $\ell=0,1,2$,
\begin{align}\label{conseq2}
2 Q^{(p)} & =\partial_t^2 Q_{kk}^{(\rho)}-2H\partial_t Q_{kk}^{(\rho)}+H \partial_t Q^{(p)}_{kk}, \\
2 Q^{(p)}_i &= (\partial_t - H)(\partial_t Q^{(\rho)}_{kki}+H(Q^{(p)}_{kki}-3 Q^{(\rho)}_{kki}))-4 S_{ki|k},\\
2Q^{(p)}_{ij} & = (\partial_t - 2H)(\partial_t Q^{(\rho)}_{kkij}+H(Q^{(p)}_{kkij}-4Q^{(\rho)}_{kkij}))-2 S_{ij|kk}-8 S_{k(i|j)k}.
\end{align}

\subsection{Consistent quadrupolar truncation} 

The quadrupolar truncation captures the dominant effects of gravitational radiation.  In all subsequent sections, we will restrict our analysis to the quadrupolar order, which is defined as fixing 
\begin{equation}
\int d^3 x a^{\ell+1} T_{\mu\nu}x^L=0,\qquad \forall \ell >2. 
\end{equation}
In the quadrupolar truncation, the conservation equations \eqref{eq37}-\eqref{eq38} imply
\begin{align}
P_{(i \vert j k )}=0, \qquad P_{i \vert j k l}=0,\qquad S_{i (j \vert kl )}= 0.
\end{align}
Eq. \eqref{27IV23.02sec} then imply $S_{(ij \vert k)}=0$. 
These conditions imply the trace conditions
\begin{equation}
P_{j\vert ji}=-\frac{1}{2}P_{i\vert jj}, \quad 
S_{ij \vert j}=-\frac{1}{2}S_{jj\vert i}=-\frac{1}{2}Q^{(p)}_{i}, \quad 
S_{ij \vert kk}=S_{kk \vert ij}=Q^{(p)}_{ij},\quad S_{i k \vert k j}= - \frac{1}{2}Q^{(p)}_{ij}. 
\end{equation}
The tensors $\epsilon_{ikl} S_{jk \vert l}$ and $\epsilon_{ilm}S_{j l\vert m k}$ 
are traceless with respect to each index contraction. Their antisymmetric parts are determined in terms of $Q^{(p)}_i$ and $Q^{(p)}_{ij}$. The symmetric tracefree part of $\epsilon_{ikl} S_{jk \vert l}$ is unconstrained. Now, the symmetric tracefree part of $\epsilon_{ilm}S_{j l\vert m k}$ reads explicitly as
\begin{equation}
\epsilon_{ilm} (S_{j l\vert m k}+S_{k l\vert m j})+ \epsilon_{jlm}(S_{k l\vert m i} +S_{i l\vert m k})+\epsilon_{klm} (S_{i l\vert m j}+ S_{j l\vert m i})=0, \label{eq:antisym S}
\end{equation}
which identically vanish after using $S_{l(i|jm)}=0$ and the fact that $S_{ij \vert kl}$ is symmetric in both $ij$ and $kl$ indices. We can then perform the following decompositions in terms of irreducible tensors  
\begin{subequations}\label{decom}
    \begin{align}
P_{i \vert jk} &=\frac{1}{2}\epsilon_{li(j}J_{k)l}-\frac{1}{2}\delta_{i(k}P_{j) \vert ll}+\frac{1}{2}\delta_{jk}P_{i \vert ll},\\
S_{i j \vert k} &= \frac{1}{2}\epsilon_{kl (i} K_{j)l}-\frac{1}{2}\delta_{k(i}Q^{(p)}_{j)}+\frac{1}{2}\delta_{ij}Q^{(p)}_k, \\
S_{i j \vert kl} &= \delta_{ij}Q^{(p)}_{kl}- (\delta_{i(k}Q^{(p)}_{l)j}+\delta_{j(k}Q^{(p)}_{l)i})+ Q^{(p)}_{ij}\delta_{kl}-\frac{1}{2} \delta_{ij}\delta_{kl}Q^{(p)}_{mm}+\frac{1}{2} \delta_{i(k}\delta_{l)j}Q^{(p)}_{mm},\label{Sijkl}
\end{align}
\end{subequations}
where we defined the symmetric and tracefree odd parity quadrupolar moments 
\begin{align}
J_{ij} &:=\frac{4}{3}P_{k \vert l (i}\epsilon_{j) kl}, \\
K_{ij}& :=\frac{4}{3}\epsilon_{kl(i} S_{j)k \vert l}. 
\end{align}
Taking the trace and the antisymmetric part in $ij$ of Eq. \eqref{27IV23.02sec}, we obtain
\begin{align}\label{consJK}
(\partial_t - H)P_{i \vert jj} &=Q_i^{(p)},\\ 
(\partial_t - H)J_{ij} &= - K_{ij}.
\end{align}

In summary we defined the multipolar moments $Q^{(p)}$, $Q^{(p)}_i$, $Q^{(p)}_{ij}$, $Q^{(\rho)}$, $Q^{(\rho)}_i$, $Q^{(\rho)}_{ij}$, $J_{ij}$, $K_{ij}$, and $P_{i \vert jj}$.
This analysis completes previous work at quadrupolar order, \text{e.g. } \cite{Ashtekar:2015lxa, Ashtekar:2015ooa, Date:2016uzr,Date:2015kma,Hoque:2018byx,Chu:2016qxp,He:2017dzb, Dobkowski-Rylko:2022dva}, which neither defined the dipole moment $P_{i \vert jj}$, the odd parity quadrupole moments $J_{ij}$ and $K_{ij}$ nor obtained the decompositions \eqref{decom} at quadrupolar order. We emphasize that conservation of the stress-energy tensor implies Eq. \eqref{Sijkl}. Imposing $S_{ij \vert kl}=0$ while keeping $Q^{(p)}_{ij} \neq 0$ as performed in various places in the literature \cite{Ashtekar:2015lxa, Ashtekar:2015ooa, Date:2016uzr,Date:2015kma,Hoque:2018byx,Chu:2016qxp,He:2017dzb, Dobkowski-Rylko:2022dva} is therefore inconsistent with the conservation of the stress-energy tensor.



\subsection{Scalar and vector inhomogeneous solutions}
\label{sec:scalar}

When matter is present, one cannot set $\hat{\chi}=0$ and $\chi_{0i}=0$ by a gauge choice for any non-vanishing source. This is the analogue of the  result in Minkowski spacetime where one cannot fix $\hat{h}_{00}=0$, $\hat{h}_{0i}=0$ inside the source. In particular, the reasoning of Section \ref{sec:gauge} fails: Eq. \eqref{res-gauge-1-} together with \eqref{11XII.01} implies 
\begin{equation}
    \Box \partial_\eta \left( \frac{\xi^0}{\eta} \right) = - \frac{4\pi \hat{T}}{\eta}.
\end{equation}
This is not consistent with \eqref{Res-Gauge-1}, the equation that residual gauge transformations need to satisfy. The argument is identical for  $\chi_{0i}$. We therefore need to solve the inhomogenous equations.

The scalar and vector inhomogeneous equations for $\hat\chi/\eta$ and $\chi_{0i}/\eta$ \eqref{11XII.01} and \eqref{11XII.02} take formally the same form as for Minkowski spacetime in coordinates $(\eta,x^i)$. Their solution in terms of the moments of the stress-tensor is therefore a consequence of the standard Minkowski Green's function. Due to the $1/\eta$ factor in the source which becomes retarded $1/(\eta - \vert \vec{x}-\vec{x}'\vert )$ in the argument of the Green function, the solution is given by
\begin{align}
\label{28II23.04}
   {\hat \chi} &= 4  \int \frac{d^3 x'}{|\vec{x}-\vec{x}'| }{} \frac{1}{1- \eta^{-1}|\vec{x}-\vec{x}'|}
\hat {T}(\eta - |\vec{x}-\vec{x}'|, {x}' ), \\
    {\chi}_{0i} &= 4  \int \frac{d^3 x'}{|\vec{x}-\vec{x}'| }{} \frac{1}{1- \eta^{-1}|\vec{x}-\vec{x}'|}
{T}_{0i}(\eta - |\vec{x}-\vec{x}'|, {x}' ). \label{28II23.04bis}
\end{align}
The solution is naturally expressed in terms of the retarded time $\eta_{\text{ret}} := \eta-\rho$. The norm $r'$ of the integration variable $\vec{x}'$ as well as $\vec{x}' \cdot \vec{n}$ are bounded by the coordinate dimension of the source $d$, at the retarded time of observation. We now assume that the sources are not of cosmological origin but are localized within the Hubble scale. More precisely, we assume that the physical size of the source $a(\eta_{\text{ret}}) d$ is smaller than the Hubble scale at any retarded time $\eta_{\text{ret}}$: $a(\eta_{\text{ret}}) d \ll H^{-1}$. This implies 
\begin{equation}
\frac{d}{\rho}= \frac{a(\eta_{\text{ret}})d}{a(\eta_{\text{ret}})\rho} \ll \frac{1}{H a(\eta_{\text{ret}}) \rho }=1-\frac{\eta}{\rho}. 
\end{equation}
Close to $\scri^+$, we have $-\eta/\rho \ll 1$. This is most easily seen in Bondi coordinates where $-\eta/\rho=(H r)^{-1}$. For  radii above the cosmological horizon we have $H r \gg 1$ which gives the desired relation $-\eta/\rho \ll 1$. Therefore, in the late time expansion close to $\scri^+$, we have both
\begin{equation}\label{scaling}
\frac{d}{\rho}\ll 1,\qquad \frac{d}{-\eta_{\text{ret}}} \ll 1. 
\end{equation}
Furthermore, we can consider sources that are slowly varying in the sense that 
\begin{equation}\label{scaling2}
T_{\mu\nu} > d\;  \partial_{\eta_{\text{ret}}} T_{\mu\nu}  > d^2\;  \partial^2_{\eta_{\text{ret}}} T_{\mu\nu} > \dots  
\end{equation}
at any fixed retarded time $\eta_{\text{ret}}$ and space $x^i$ inside the source. In summary, the scaling \eqref{scaling} applies close to $\scri^+$ for any sources below the Hubble scale and the approximation \eqref{scaling2} applies for a slowly moving source. These relations are identical to the ones derived in \cite{Ashtekar:2015lxa}. 

We can now solve Eqs. \eqref{28II23.04}-\eqref{28II23.04bis} in the quadrupolar approximation as follows. We first expand 
\begin{align}\label{xxp}
|\vec{x}-\vec{x}'|
&= \rho\bigg(1 - \frac{\vec{n} \cdot \vec{x}'}{\rho} + \frac{r'^2-(\vec{n} \cdot \vec{x}')^2}{2\rho^2}+O(\frac{d^3}{\rho^3})\bigg),\\
|\vec{x}-\vec{x}'|^{-1}
=&\frac{1}{\rho}\bigg(1 +\frac{\vec{n} \cdot \vec{x}'}{\rho} - \frac{r'^2-3(\vec{n} \cdot \vec{x}')^2}{2\rho^2}+O(\frac{d^3}{\rho^3})\bigg), \\ 
\frac{\eta}{\eta - |\vec{x}-\vec{x}'|} &=  \frac{\eta}{\eta -\rho} \bigg(1 -\frac{\vec{n}\cdot \vec{x}'}{\eta -\rho}+ \frac{(\vec{n}\cdot \vec{x}')^{2}}{(\eta -\rho)^{2}}  + \frac{r'^2 -(\vec{n}\cdot \vec{x}')^{2}}{2 \rho (\eta -\rho)} +O(\frac{d}{\rho}\frac{d^2}{(-\eta_{\text{ret}}^2)})\nonumber \\ 
&+O(\frac{d^2}{\rho^2}\frac{d}{(-\eta_{\text{ret}})})+O(\frac{d^3}{(-\eta_{\text{ret}})^3})\bigg).
\end{align}
We substitute $\eta = \eta_{\text{ret}}+\rho$ in the above expansion as well as in the argument of $\hat T$ and $T_{0i}$ and expand up to second order in powers of $d/\rho$ and $d/(-\eta_{\text{ret}})$. By construction, this leads to the quadrupole approximation which keeps all terms at most quadratic in $r'$ and $n \cdot \vec{x}'$. The expansion is justified close to $\scri^+$ for slowly moving sources because of the scaling \eqref{scaling} and the approximation \eqref{scaling2}. Far from $\scri^+$ or for arbitrary moving sources higher multipoles will also contribute. In this paper we truncate to the quadrupole order. The inclusion of higher multipoles is beyond the scope of this paper. After some algebra we obtain 
\begin{eqnarray} \nonumber
\hat{\chi} &=& \frac{4 \eta }{\rho(\eta-\rho)} \int d^{3}x'  
\bigg[ \hat{T} 
(\eta_{\text{ret}}, x')+\hat{T}(\eta_{\text{ret}}, x')\bigg( -\frac{\vec{n}\cdot 
\vec{x}'}{\eta -\rho}+ \frac{(\vec{n}\cdot 
\vec{x}')^{2}}{(\eta -\rho)^{2}}\bigg)+  
\hat{T}^{(1)} (\eta_{\text{ret}}, x') (\vec{n}\cdot 
\vec{x}') \\ \nonumber
&& \times \bigg( 1 -\frac{\vec{n}\cdot \vec{x}'}{\eta 
-\rho}\bigg)+\frac {1}{2} (\vec{n} \cdot \vec{x}')^{2} 
\hat{T}^{(2)} (\eta_{\text{ret}}, x')
+ \frac{1}{\rho} \bigg( \hat{T} (\eta_{\text{ret}}, x')\bigg(\vec{n}\cdot \vec{x}'- \frac{3 (\vec{n}\cdot x')^{2} - r'^{2}}{ 2(\eta - \rho)}\bigg) 
\\ \label{12VI23.02}
&&+ \hat{T}^{(1)}(\eta_{\text{ret}}, x')
\frac{3(\vec{n}\cdot \vec{x})^{2}- r'^{2}}{2}\bigg)+
\frac{1}{\rho^{2}} \bigg( \hat{T}(\eta_{\text{ret}}, x') \frac{3(\vec{n}\cdot \vec{x})^{2}- r'^{2}}{2}\bigg)
\bigg].
\end{eqnarray}
The corresponding expression for $\chi_{0i}$ can be formally obtained from the expression for $\hat \chi$ after replacing $\hat T$ by $T_{0i}$.

We now substitute $(\eta-\rho)^{-1}= \eta_{\text{ret}}^{-1}=-H a(\eta_{\text{ret}})$ and $\frac{\eta}{\eta-\rho}=1-H\rho a(\eta_{\text{ret}})$. Using the definition of the moments of the stress-energy tensor and their conservation equations up to the quadrupolar order, we have 
\bea
\int d^{3}x' a (\eta) \hat{T} (\eta, x')&=&\int d^{3}x' a (\eta)(T_{00}+\delta^{ij}T_{ij})=Q^{(\rho+p)},\nonumber\\
\int d^{3} x' a (\eta) x'_{k} \partial_{\eta} \hat T&=&\int d^{3} x' a^{2} (t) x'_{k} \partial_{t} \hat T= \partial_{t} Q^{(\rho+p)}_{k} -2HQ^{(\rho+p)}_{k},\\
 \int d^{3} x' a (\eta) x'_{k} x'_{l} \partial_{\eta}^{2} \hat T &=& \int d^{3} x' a^{3}(t) x'_{k} x'_{l} (\partial_{t}^{2}+H\partial_{t}) \hat T =\partial_{t}^{2} Q^{(\rho+p)}_{kl}-5H\partial_{t} Q^{(\rho+p)}_{kl}+6H^{2}Q^{(\rho+p)}_{kl}.\nonumber
\eea
and 
\bea 
\int d^{3}x' a(\eta)x'_{k} T^{(1)}_{0i}(\eta) 
&=& \int d^{3}x' a^{2}(t) x'_{k}\partial_{t}
T_{0i} (\eta, x') 
= \partial_{t} P_{i|k} (\eta) - 2H P_{i|k} (\eta)  ,\\
 \int d^{3} x' a (\eta) x'_{k} x'_{l} \partial_{\eta}^{2} T_{0i} &=& \int d^{3} x' a^{3}(t) x'_{k} x'_{l} (\partial_{t}^{2}+H\partial_{t}) T_{0i}=\partial_t^2 P_{i \vert kl}-5H \partial_t P_{i \vert kl}+6H^2 P_{i \vert kl}.\nonumber
\eea
Writing \eqref{12VI23.02} in terms of the source moments, we obtain
\bea \nonumber
\hat{\chi} (\eta, x)&=& 4 \bigg(\frac{1}{ a \rho}-H \bigg) \bigg(Q^{(\rho+p)}  - \frac{H}{2} (\partial_{t} -H)Q_{kk}^{(\rho+p)}+ n_{k}\partial_{t} Q_{k}^{(\rho+p)} +\frac{n_{k}n_{l}}{2} (\partial_{t}^{2} -H^{2})Q_{kl}^{(\rho+p)} 
 \bigg)
\\ 
&& 
\hspace{-2cm}
+
4 \bigg(\frac{1}{ a \rho}-H \bigg)^{2} 
\bigg( n_{k} Q_{k}^{(\rho+p)}+\frac{1}{2}(3n_{k}n_{l}-\delta_{kl}) \partial_{t} Q_{kl}^{(\rho+p)}
\bigg)+2
\bigg(\frac{1}{ a \rho}-H \bigg)^{3} (3n_{k}n_{l}-\delta_{kl}) Q_{kl}^{(\rho+p)}\! ,\nonumber\\ \label{finalchi1}
\eea
where the expression is evaluated at $\eta_{\text{ret}}$. The expression for $\chi_{0i}$ is exactly analogous to the expression for $\hat\chi$ with the replacement $Q^{(\rho+p)}\mapsto P_i$, $Q^{(\rho+p)}_j\mapsto P_{i \vert j}$, $Q^{(\rho+p)}_{jk}\mapsto P_{i \vert jk}$. We now use the conservation equations truncated at quadrupolar order to rewrite $\chi_{0i}$ as
\bea \nonumber
\chi_{0i}(\eta, x) &=& 4 \bigg(\frac{1}{ a \rho}-H \bigg) \bigg(P_{i} -\frac{H}{2}(\partial_{t}-H)P_{i|ll}-n_{k} S_{ik}+\frac{n_k n_l}{2}(\partial_{t}+ H)S_{kl|i}\bigg)
\\  \nonumber
&& 
\hspace{-2.5cm}
+
4 \bigg(\frac{1}{ a \rho}-H \bigg)^{2} 
\bigg( n_k P_{i|k}-\frac{1}{2}\partial_{t} P_{i|ll}+\frac{3 n_k n_l}{2}(S_{kl|i}+HP_{i|kl})\bigg)+2
\bigg(\frac{1}{ a \rho}-H \bigg)^{3} (3n_{k}n_{l}-\delta_{kl}) P_{i|kl} ,\\
\label{finalchi0i}\eea
evaluated at $\eta_{\text{ret}}$.

We emphasize that these expressions are exact to all orders in $\rho$ in the quadrupole approximation. Note that in Bondi coordinates, $\frac{1}{a(\eta_{\text{ret}})\rho}=H+\frac{1}{r}$ as we will see in Section \ref{sec:Bondi}.

\subsection{Tensor inhomogeneous solution}
\label{sec:tensor}

We now turn to the inhomogeneous equation for $\chi_{ij}$ \eqref{decoupled3} which can be alternatively written as 
\[
\Box \chi_{ij} + \frac{2}{\eta}\partial_0 \chi_{ij} = - 16\pi T_{ij}.
%
\]
The corresponding retarded Green function is defined as 
\begin{equation}
\left(\Box + \frac{2}{\eta}\partial_0\right) G_R(\eta, x; \eta', x') = -
\frac{\Lambda}{3}\eta^2\delta^4(x - x') ~ ,
\end{equation}
and is given by  \cite{Ford:1977dj,Waylen:1978dx,Caldwell:1993xw,deVega:1998ia} 
\begin{equation} \label{NonCovariantGreenFn}
G_R(\eta, x; \eta' x') ~ = ~ \frac{\Lambda}{3}\eta \eta' \frac{1}{4\pi}
\frac{ \delta(\eta - \eta' - |x - x'|) }{|x - x'|} ~+~
\frac{\Lambda}{3}\frac{1}{4\pi}\theta(\eta - \eta' - |x - x'|) ~ .
\end{equation}
The theta function leads to cosmological tails induced by the background curvature. The particular solution is given by
\begin{eqnarray}
\chi_{ij}(\eta, x)\!\! & = & 16\pi \int_{\mathrm{source}} \frac{d\eta'
d^3x'}{\frac{\Lambda}{3}\eta'^2} G_R(\eta, x; \eta' x') T_{ij}(\eta',
x') \\
& & \hspace{-2cm} = 4\int d\eta'd^3x' \frac{\eta}{\eta'}\frac{\delta(\eta - \eta' - |x
- x'|) }{|x - x'|} T_{ij}(\eta', x') \nonumber \\
& & \hspace{1.0cm} ~ + ~ 4\int d\eta'd^3x' \frac{1}{\eta'^2}\theta(\eta
- \eta' - |x - x'|) T_{ij}(\eta', x') \label{RadiativeSoln0} \\
%
%
& & \hspace{-2cm} =4 \int d^3x' \frac{\eta}{|x - x'|(\eta - |x - x'|)} \left.
T_{ij}(\eta', x') \right|_{\eta' = \eta - |x - x'|} 
\nonumber 
%
+ ~ 4 \int d^3x' \int_{- \infty}^{\eta - |x
-x'|}d\eta'\frac{T_{ij}(\eta', x')}{\eta'^2} \\ \nonumber
& & \hspace{-2cm}= 4 \int  \frac{d^3x'}{|x - x'|} \left.
T_{ij}(\eta', x') \right|_{\eta' = \eta - |x - x'|} 
%
+ ~ 4 \int d^3x' \int_{- \infty}^{\eta - |x
-x'|} \frac{d\eta'}{\eta'} \partial_{\eta'}{T_{ij}(\eta', x')} \label{30X22.02} ,
\label{RadiativeSoln}
\end{eqnarray}
where in the last line we used that $T_{ij}(-\infty,x^i)$ is finite. We further split the tail integral in two parts: between $-\infty$ and $\eta_{\text{ret}}:=\eta-\rho$ and between $\eta_{\text{ret}}$ and $\eta-|x - x'|$. We shift the integration variable of the second integral $\eta' \mapsto \eta'+\eta_{\text{ret}}$ to obtain
\begin{align}
\chi_{ij}(\eta,x)&=4\int \frac{d^3 x'}{|x - x'|}T_{ij}(\eta-|x - x'|,x')+4\int d^3 x' \int_{-\infty}^{\eta_{\text{ret}}} \frac{d\eta'}{\eta'} \partial_{\eta'}{T_{ij}(\eta', x')}\nonumber \\ 
&+\frac{4}{\eta_{\text{ret}}} \int d^3 x' \int_0^{\rho -|x - x'| } \frac{d\eta'}{1+\frac{\eta'}{\eta_{\text{ret}}}}  \partial_{\eta'}{T_{ij}(\eta_{\text{ret}}+\eta', x')} \label{chiijIN}\\
&= \chi^{(I)}_{ij}(\eta,x)+\chi^{(II)}_{ij}(\eta,x)+\chi^{(III)}_{ij}(\eta,x).
\end{align}

In order to simplify the second term $\chi^{(II)}_{ij}$ -- which we call the tail term -- we interchange the $\vec{x}'$ and $\eta'$ integrals, we switch the integration variable to cosmological time $t$ defined as $\eta=-H^{-1}e^{-Ht}$ and we use the conservation equation \eqref{27IV23.02first} which can be written as $\int d^3 x'T_{ij}(\eta,x')=-a(\eta)^{-1}\partial_t P_{(i \vert j)}(\eta)$ where $P_{(i \vert j)}=-\frac{1}{2}(\partial_t Q^{(\rho)}_{ij}-2H Q^{(\rho)}_{ij}+H Q^{(p)}_{ij})$. The tail term becomes 
\begin{align}
\chi^{(II)}_{ij} &= 4 H \int_{-\infty}^{t_{\text{ret}}} dt e^{H t}\partial_t \left( e^{-H t} \partial_t P_{i \vert j}(\eta) \right) = 4 H \int_{-\infty}^{t_{\text{ret}}} dt \partial_t \left( (\partial_t - H) P_{i \vert j}(\eta) \right),
\end{align}
which is a total derivative. It is remarkable that using the conservation equation, one can reduce this tail integral as a difference between boundary terms. This step is new with respect to the treatment of \cite{Ashtekar:2015lxa}. Substituting the expression for $P_{(i \vert j)}$, we find the exact expression for the tail integral 
\begin{align}
\chi_{ij}^{(II)}& =-4H \bigg[S_{ij}(\eta)+HP_{(i|j)}(\eta)\bigg]^{t_{\text{ret}}}_{-\infty} \\
&=-2H (\partial_{t}^{2} Q_{ij}^{(\rho)}-3H\partial_{t}Q_{ij}^{(\rho)}+2H^2 Q_{ij}^{(\rho)}+H\partial_{t}Q_{ij}^{(p)}-H^{2}Q_{ij}^{(p)})(\eta_{\text{ret}})+\chi_{ij}(-\infty) ,\label{10IX22.1f}
\end{align}
where 
\bea \label{chiijinf} 
\chi_{ij}(-\infty):=2H (\partial_{t}^{2} Q_{ij}^{(\rho)}-3H\partial_{t}Q_{ij}^{(\rho)}+2H^2 Q_{ij}^{(\rho)}+H\partial_{t}Q_{ij}^{(p)}-H^{2}Q_{ij}^{(p)})(-\infty)
\eea
is the only non-instantaneous term. 

We will now expand the first and third terms in Eq. \eqref{chiijIN} in the multipolar expansion. Expanding $|x - x'|$ as Eq. \eqref{xxp} and Taylor expanding we have schematically
\begin{align}
\chi^{(I)}_{ij} &=4 \int d^3 x' \Bigg( \frac{ T_{ij}(\eta_{\text{ret}},x') + \vec n \cdot \vec{x}'\, T^{(1)}_{ij}(\eta_{\text{ret}},x')+\frac{1}{2} (\vec n \cdot \vec{x}')^2 T^{(2)}_{ij}(\eta_{\text{ret}},x')+\dots }{\rho}\nonumber \\
&+\frac{ \vec n \cdot \vec{x}'\, T_{ij}(\eta_{\text{ret}},x')+\frac{1}{2}( 3 (\vec n \cdot \vec{x}')^2-r'^2)T^{(1)}_{ij}(\eta_{\text{ret}},x')+\dots }{\rho^2}\nonumber \\
&+\frac{(3 (\vec n \cdot \vec{x}')^2-r'^2)\, T_{ij}(\eta_{\text{ret}},x')+\dots }{2\rho^3}+O(\rho^{-4}) \Bigg) \label{expij} .
\end{align}
In the late-time approximation and in the slow motion approximation of the source, we can use Eqs. \eqref{scaling} and \eqref{scaling2}. In the quadrupolar truncation, only the displayed terms in Eq. \eqref{expij} are non-vanishing. Note that our definition of quadrupolar truncation is more general than \cite{Ashtekar:2015lxa} where fewer terms are kept. Using the definitions of the moments of the stress-tensor we obtain 
\begin{align}
\chi_{ij}^{(I)}  &=  \frac{4} {a(\eta_{\text{ret}}) \rho} \Bigg( S_{ij}+n_k(\partial_t S_{ij \vert k}-2 H S_{ij \vert k})+\frac{1}{2}n_kn_l (\partial_t^2 S_{ij\vert kl}-5H \partial_t S_{ij\vert kl}+6H^2 S_{ij\vert kl} )\nonumber\\
 &+\frac{1}{\rho a(\eta_{\text{ret}})}\bigg( n_k S_{ij\vert k} + \frac{1}{2 }(3n_kn_l-\delta_{kl})(\partial_t S_{ij \vert kl}-3 H S_{ij \vert kl})\bigg)+\frac{3n_kn_l-\delta_{kl}}{2 a^2(\eta_{\text{ret}})\rho^2} S_{ij \vert kl}\Bigg),
\end{align}
where all terms are evaluated at $\eta_{\text{ret}}$. 

In order to evaluate the third integral we note that $\delta := \rho - |x - x'|$ is $O(d)$. Expanding up to second order in $d$ we obtain 
\begin{align}
\int_0^{\delta} \frac{d\eta'}{1+\frac{\eta'}{\eta_{\text{ret}}}}  \partial_{\eta'}{T_{ij}(\eta_{\text{ret}}+\eta', x')}= T^{(1)}_{ij}(\eta_{\text{ret}}, x')\delta +\left(T^{(2)}_{ij}(\eta_{\text{ret}}, x')-\frac{1}{\eta_{\text{ret}}}T^{(1)}_{ij}(\eta_{\text{ret}}, x') \right)\frac{\delta^2}{2}. 
\end{align}
Using Eq. \eqref{xxp}, we have 
\begin{align}
\chi_{ij}^{(III)} &=  \int d^3x' \Bigg( 4(\vec{n}\cdot \vec{x}'-\frac{r^{\prime 2}-(\vec{n}\cdot \vec{x}')^2}{2\rho})\frac{T_{ij}^{(1)}(\eta_{\text{ret}},x')}{\eta_{\text{ret}}}\nonumber\\
&\qquad +\frac{2 (\vec{n}\cdot \vec{x}')^2}{\eta_{\text{ret}}}\left(T_{ij}^{(2)}(\eta_{\text{ret}},x') -\frac{T_{ij}^{(1)}(\eta_{\text{ret}},x')}{\eta_{\text{ret}}}\right)\Bigg)\\
&= -4H \Bigg[n_k (\partial_t S_{ij \vert k}-2H S_{ij \vert k}) -\frac{1}{2\rho a(\eta_{\text{ret}})}(\delta_{kl}-n_k n_l)(\partial_t S_{ij \vert kl}-3HS_{ij \vert kl})\nonumber \\ 
&+\frac{n_k n_l}{2} \bigg( \partial_t^2 S_{ij \vert kl}-4 H \partial_t S_{ij \vert kl}+3H^2 S_{ij \vert kl} \bigg) \Bigg].
\end{align}
In total, we obtain the metric components in terms of irreducible multipole moments:
\begin{align} \nonumber
\chi_{ij}^{(I)}+\chi_{ij}^{(II)}+\chi_{ij}^{(III)} &=\chi_{ij}(-\infty)+4H^{2}\bigg(-P_{(i|j)}+n_{k}S_{ij|k}+\frac{1}{2} n_{k}n_{l} \partial_{t}S_{ij|kl}-\frac{H}{2}S_{ij|kk}\bigg)\\
\nonumber
& \hspace{-3cm}+4\bigg(\frac{1}{a \rho}-H\bigg)\bigg( S_{ij}+n_k\partial_t S_{ij \vert k}+\frac{n_k n_l}{2} \partial_{t}^{2} S_{ij|kl}-\frac{H}{2} \partial_{t} S_{ij|kk}\bigg)\nonumber \\
&\hspace{-3cm} +4\bigg(\frac{1}{a \rho}-H\bigg)^{2}\bigg(n_{k}S_{ij|k}+\frac{1}{2}(3n_{k}n_{l}-\delta_{kl})\partial_{t} S_{ij|kl}\bigg)
+2 \bigg(\frac{1}{a\rho} -H\bigg)^{3} (3n_{k}n_{l}-\delta_{kl}) S_{ij|kl}
,\label{finalchijbis}
\end{align}
where we wrote $a := a(\eta_{\text{ret}})$ for short. We can write this expression in a simpler fashion by introducing 
\begin{equation}
\chi_{ij}^{(0)}(\eta_{\text{ret}},x^A):= \chi_{ij}(-\infty)+4H^{2} \left. \bigg(-P_{(i|j)}+n_{k}S_{ij|k}+\frac{1}{2} n_{k}n_{l} \partial_{t}S_{ij|kl}-\frac{H}{2}S_{ij|kk}\bigg) \right|_{\eta_{\text{ret}}}\label{chi0} ,
\end{equation}
where $\chi_{ij}(-\infty)$ is defined in Eq. \eqref{chiijinf}. One has the conservation law
\begin{equation}\label{conservationchij0}
\partial_t \chi_{ij}^{(0)} = 4H^2 \left( S_{ij}+n_k \partial_t S_{ij \vert k}+\frac{1}{2}n_k n_l \partial_t^2 S_{ij \vert kl}-\frac{H}{2}\partial_t S_{ij \vert kk}\right). 
\end{equation}
The metric perturbation then takes the final form 
\begin{align} \nonumber
\chi_{ij} &=\chi_{ij}^{(0)} +\bigg(\frac{1}{a \rho}-H\bigg)H^{-2}\partial_t \chi_{ij}^{(0)}  +4\bigg(\frac{1}{a \rho}-H\bigg)^{2}\bigg(n_{k}S_{ij|k}+\frac{1}{2}(3n_{k}n_{l}-\delta_{kl})\partial_{t} S_{ij|kl}\bigg)
\nonumber\\
&+2 \bigg(\frac{1}{a\rho} -H\bigg)^{3} (3n_{k}n_{l}-\delta_{kl}) S_{ij|kl}
.\label{finalchij}
\end{align}
The metric is most naturally expanded in terms of the Bondi-Sachs radius expansion $\frac{1}{r} = \frac{1}{a\rho}-H+O(h)$, as will be described in Section \ref{sec:Bondi}. Note that $H^{-2}\chi_{ij}^{(0)}$ does admit a flat limit since $\chi_{ij}^{(0)}$ is proportional to $H^2$. 

Finally, using the decomposition \eqref{decom} and the conservation equations, one can write the tensorial perturbation in terms of irreducible tensors. We can write $\chi_{ij}^{(0)}$  as
\begin{align} \label{chi0eq}
\chi_{ij}^{(0)}(\eta_{\text{ret}},x^A)&= \chi_{ij}(-\infty) +4H^{2} \bigg(\frac{1}{2} \partial_{t}Q_{ij}^{(\rho+p)}-HQ_{ij}^{(\rho)}+\frac{1}{2}n_{k}\epsilon_{kl(i}K_{j)l}-\frac{1}{2} n_{(i}Q_{j)}^{(p)}+\frac{1}{2} \delta_{ij} n_kQ_{k}^{(p)}\nonumber \\ 
& \left. +\frac{\delta_{ij}}{2} n_{k} n_{l} \partial_{t}Q_{kl} ^{(p)} - n_{l} n_{(i}\partial_{t}Q_{j)l}^{(p)} -\frac{1}{4} (\delta_{ij}-n_{i}n_{j}) \partial_{t}Q_{mm}^{(p)}
\bigg) \right|_{\eta_{\text{ret}}}.
\end{align}
Note that the $\frac{H}{2} Q_{ij}^{(p)}$ term originating from $-P_{(i\vert j)}$ after using the 
conservation equation \eqref{Pijd} remarkably cancels 
out with the term $-\frac{H}{2} Q_{ij}^{(p)}$ originating from $-\frac{H}{2} S_{ij|kk}$ after using the structural equation \eqref{Sijkl} of the higher order tensor $S_{ij \vert kl}$. 
\footnote{This implies that terms proportional to $Q_{ij}^{(p)}$ in Eq. (3.22) of \cite{Ashtekar:2015lxa}, and in Eq. (96) of \cite{Date:2015kma} exactly cancel out once one also includes the higher order contributions of the Taylor expansion of the source integral as we performed in Eq. \eqref{expij}.}

Following the $tt$ notation of \cite{Ashtekar:2017ydh}, the $tt$ transverse-traceless projection of $\chi_{ij}$ is given by 
\begin{align}
\chi_{mn}^{tt}&:= \perp^{tt}_{ij|mn} \chi_{ij}= \chi_{mn}^{tt}(-\infty)+4H^{2}\bigg(\frac{1}{2} \partial_{t}Q_{ij}^{(\rho+p)}-H Q_{ij}^{(\rho)}+\frac{1}{2}n_{k}\epsilon_{kp(i}K_{j)p}\bigg)\perp^{tt}_{ij|mn}\nonumber \\ \nonumber
&+4\bigg(\frac{1}{a\rho}-H\bigg)\partial_{t}\bigg(\frac{1}{2} \partial_{t}Q_{ij}^{(\rho+p)}-H Q_{ij}^{(\rho)}+\frac{1}{2}n_{k}\epsilon_{kp(i}K_{j)p}\bigg) \perp^{tt}_{ij|mn} \\
&+4 \bigg(\frac{1}{a\rho}-H\bigg)^{2} \bigg(\partial_{t}Q_{ij}^{(p)}+\frac{1}{2}n_{k} \epsilon_{kp(i}K_{j)p}\bigg)\perp^{tt}_{ij|mn}
+4 \bigg(\frac{1}{a\rho}-H\bigg)^{3} Q_{ij}^{(p)} \perp^{tt}_{ij|mn},
\end{align}
where $\perp^{tt}_{ij|kl}:=\perp_{k(i}\perp_{j)l}-\frac{1}{2} \perp_{ij}\perp_{kl}$ and $\perp_{ij}:=\delta_{ij}-n_i n_j$. 
Note that $\chi_{ij}^{tt}$ does not depend upon $P_{i \vert kk}$, $Q^{(p)}$ nor $Q^{(p)}_i$. For a stationary source, the coefficient in front of $(\frac{1}{a\rho}-H)$ (which is $1/r$ in Bondi-Sachs coordinates) vanishes, while the other terms do not vanish in general.

\subsection{Flat limit}
\label{sec:flat}

As a check, let us now obtain the standard $\Lambda=0$ limit of these expressions. We define the moments of the stress-energy tensor as the flat limit of Eqs. \eqref{25IV23.02} and \eqref{25IV23.02bis},
\bea
Q_L(t) &:= &\int  T_{00} (t, x') x'_L\, d^{3}x', \\
P_{i|L}(t) &:= &\int  T_{0i} (t, x') x'_L\, d^{3}x',
\eea
Starting from harmonic gauge, one needs in general to perform a coordinate transformation to reach canonical harmonic gauge \cite{Thorne:1980ru,Damour:1990gj}. This coordinate transformation is avoided if the multipolar content of the metric perturbation is restricted. For simplicity of this check here, we restrict the multipolar content by imposing 
\begin{equation}
Q^{(p)}=0,\qquad Q_i^{(p)}=0,\qquad Q_{ij}^{(p)}=0,\qquad P_{i \vert kk}=0. \label{restr}
\end{equation}
(In case we do not impose this restriction, a transformation to canonical harmonic gauge would be required, which we do not perform here.)

The flat limit $H \to 0$ leads to set
\bea 
\qquad Q_{L}^{(\rho)} \to M_{L},\qquad  S_{ij}\to \frac{1}{2}\ddot M_{ij}, \qquad Q^{(p)}\to \frac{\delta^{ij}}{2} \ddot M_{ij},\qquad P_{i\vert j}\mapsto \frac{1}{2}\epsilon_{ijk}J_k-\frac{1}{2}\dot M_{ij}.
\eea
We furthermore switch to the standard flat definition of current multipole moments $S_L := \frac{\ell+1}{2\ell}J_L$ to express the solution in terms of the mass and current canonical multipole moments $M_L$ and $S_L$. 

According to the map \eqref{map}, the trace-reversed metric perturbations $\hat h_{\mu\nu}$ are given in terms of $\hat\chi$, $\chi_{0i}$, $\chi_{ij}$ as
\begin{equation}
\hat h_{00}=a^2 \hat\chi-a^2\chi_{ii}, \qquad \hat h_{0i}=a^2 \chi_{0i},\qquad \hat h_{ij}=a^2\chi_{ij}. 
\end{equation}
Taking the flat limit of Eqs. \eqref{finalchi1}, \eqref{finalchi0i} and \eqref{finalchij} and restricting as Eq. \eqref{restr}, we recover the standard trace-reversed linearized perturbations up to the quadrupolar order in canonical harmonic gauge 
\begin{align}
\hat{h}_{00}(t,\vec{x}) &= \frac{4M-4 n_i \dot P_i+2  n_i  n_j \ddot M_{ij}}{\rho}+\frac{4 n_i M_i + (6 n_i  n_j-2\delta_{ij})\dot M_{ij}}{\rho^2} +\frac{(6 n_i  n_j-2\delta_{ij})M_{ij}}{\rho^3}, \\ \label{2VII23.01}
\hat{h}_{0i}(t,\vec{x}) &= \frac{4P_{i}-2  n_j \ddot M_{i j}+\frac{4}{3}\epsilon_{ilj}n_{k}n_{l} \ddot S_{kj}}{\rho}+\frac{2 \epsilon_{ikl}n_{k}S_{l}-2n_{k}\dot{M}_{ik}+4\epsilon_{ijl}n_{k}n_{l}\dot S_{kl}}{\rho^2}+\frac{4\epsilon_{ikl}n_{j}n_{k}S_{jl}}{\rho^{3}} , \\
\hat{h}_{ij}(t,\vec{x}) &= \frac{2\ddot M_{ij}-\frac{8}{3}\epsilon_{kl(i}\ddot S_{j)l} n_{k}}{\rho}-\frac{\frac{8}{3}\epsilon_{kl(i}\dot S_{j)l}n_{k}}{\rho^{2}}.  
\end{align}
This exactly matches with Eq. (3a) of \cite{Blanchet:2020ngx} after changing the convention $P_i \mapsto -P_i$ for the momentum (because it is defined from $T^{0i}$ with one upper $0$ index in \cite{Blanchet:2020ngx} but one lower $0$ index here) and identifying our $\hat h_{\mu\nu}$ as $-\eta_{\mu\alpha}\eta_{\nu\beta}h_1^{\alpha\beta}$ where $h_1^{\alpha\beta}$ is defined in Eq. (3a) of \cite{Blanchet:2020ngx}.

\section{Quadrupolar perturbation in Bondi gauge}
\label{sec:Bondi}

We obtained the linearized solution up to quadrupolar order in Sections \ref{sec:scalar} and \ref{sec:tensor} in generalized harmonic gauge. We will now derive the Bondi-Sachs form of this solution. 

\subsection{Bondi gauge and $\Lambda-$BMS 
 symmetries}
 \label{sec:BMS}

\begin{figure}
    \centering
    ~~~~~
        \includegraphics[width=0.5\textwidth]{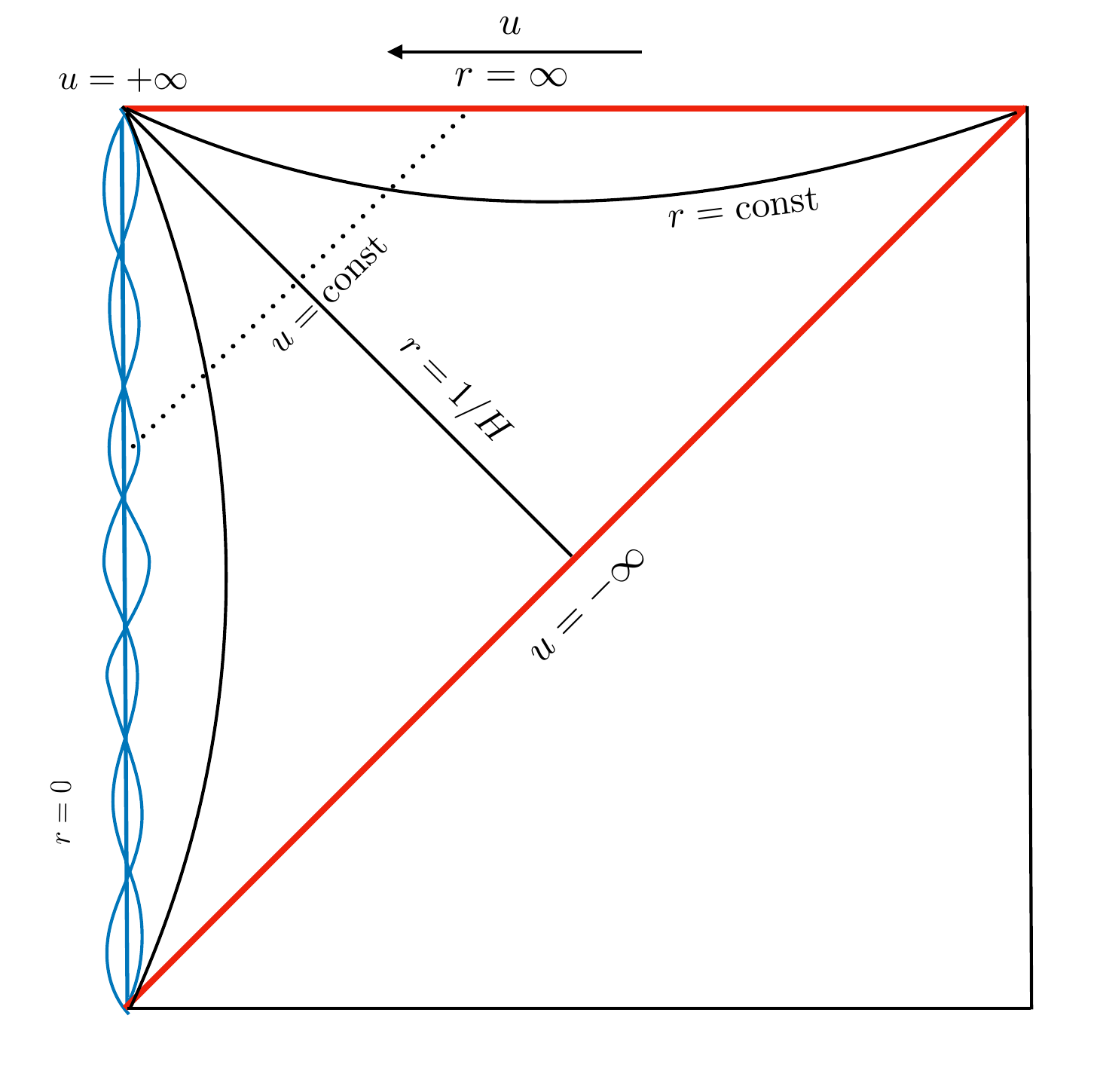}
    \qquad\qquad 
    \caption{Penrose diagram of de Sitter spacetimes in Bondi-Sachs coordinates. Blue wiggly lines denote sources that are emitting gravitational radiation. Bondi-Sachs coordinates cover the future Poincar\'e patch.}\label{PoincarePatchBondi}
\end{figure}

Bondi-Sachs coordinates $X^\mu := (u, r, x^{A})$ are adapted to outgoing null hypersurfaces $u$ constant. The normal covector to such hypersurfaces is null, i.e. $g^{\mu \nu} (\partial_{\mu}u) (\partial_{\nu}u)=0$, so that $g^{uu}=0$. The two angular coordinates $x^{A} (\theta, \phi)$, are constant along null rays, i.e. $g^{\mu\nu} (\partial_{\mu}u)(\partial_{\nu} x^{A})=0$, so that $g^{uA}=0$. Using the transformation between contravariant and covariant components of the metric, the condition $g^{uu}=0=g^{uA}$ is equivalent to  $g_{rr}=0=g_{rA}$. The radius is chosen to be the luminosity distance, which amounts to fixing $\partial_r \text{det}(g_{AB}/r^2)=0$. In Bondi-Sachs gauge, the metric can be written as 
\bea 
g_{\mu\nu} dX^{\mu} dX^{\nu}=\frac{V}{r} e^{2\beta} du^{2}-2e^{2\beta} du dr+r^{2} \gamma_{AB} (dx^{A} -U^{A}du) (dx^{B} -U^{B}du).
\eea
The explicit general expansion of an asymptotically (locally) de Sitter spacetime was derived in \cite{Compere:2019bua} (see \cite{Poole:2018koa} in the case of axial symmetry and \cite{Chrusciel:2021ttc,Chrusciel:2020rlz, Bonga:2023eml} for subsequent work). The most general class of asymptotically (locally) de Sitter metrics takes the form
\begin{align}
g_{AB} &=r^2 q_{AB}(u,x^B)+O(r), \\
\beta&=\beta_0(u,x^B)+O(r^{-2}), \\
U^A &= U^A_0(u,x^B) + O(r^{-1}), \\
\frac{V}{r}&= \frac{\Lambda}{3}e^{2\beta_{0}}r^2+O(r). 
\end{align}
The ``holographic three-dimensional boundary metric'' $g_{ab}^{(0)}$, $x^a := (u,x^A)$, is defined from its components
\begin{eqnarray}
g_{uu}^{(0)} &=& \frac{\Lambda}{3}e^{4\beta_0}+U_0^A U_A^0, \\
g_{uA}^{(0)}&=&-U^0_A, \\
g^{(0)}_{AB} &=& q_{AB}. 
\end{eqnarray}
Following \cite{Compere:2019bua}, we can perform the following ``boundary gauge fixing'' in the addition to the Bondi-Sachs gauge fixing: 
\begin{align}\label{boundarygauge}
U^A_0=0,\qquad \beta_0 = 0,\qquad \text{det}(q_{AB}) = \text{det}(\mathring{q}_{AB}),
\end{align}
where $\mathring{q}_{AB}$ is the metric on the unit 2-sphere. We will denote as $\mathring{D}_A$ its associated covariant derivative. In this boundary gauge, the coordinate $u$ is a Gaussian normal spatial coordinate and the holographic metric reads as 
\begin{align}
g_{ab}^{(0)}dx^a dx^b=H^2 du^2+ q_{AB}(u,x^C)dx^A dx^B  ,\qquad \text{det}(q_{AB})=  \text{det}(\mathring{q}_{AB}).\label{boundarygaugeresult}
\end{align}
This boundary gauge fixing can always be performed locally without loss of generality. The residual gauge transformations form the $\Lambda$-BMS algebroid whose structure constants depend upon $q_{AB}$ \cite{Compere:2019bua,Compere:2020lrt}. In the presence of radiation, an observer located close to $\scri^+$ cannot gauge fix the diffeomorphism group further. The $\Lambda$-BMS symmetries are the analogue in de Sitter space of the BMS symmetries of asymptotically flat spacetime: they reflect the freedom at setting up a detector in the presence of radiation at $\scri^+$. For this reason, we will sometimes call Bondi gauge together with the boundary gauge fixing \eqref{boundarygaugeresult} as $\Lambda$-BMS gauge.

In the absence of radiation where $q_{AB}$ reduces to the unit metric on the round sphere $\mathring q_{AB}$, the $\Lambda$-BMS groupoid become an ordinary group, which we call the $\Lambda$-BMS group. It contains but extends the $SO(1,4)$ exact symmetry group of de Sitter. The generators admit the expansion $\xi^u := \mathring{\xi}^u(u,x^A)$ and $\xi^A := \mathring{\xi}^A(u,x^A)+O(r^{-1}) $ where the leading order generators obey the following differential equations 
\begin{eqnarray}\label{Eqsxi}
\partial_u \mathring\xi^u-\frac{1}{2}\mathring{D}_A \mathring\xi^A =0, \qquad 
\partial_u \mathring\xi^A+H^2 \mathring{q}^{AB}\partial_B \mathring\xi^u = 0. 
\end{eqnarray}
The general solution is given by 
\cite{Compere:2020lrt}
\begin{subequations}\label{BMSel}
\begin{eqnarray}
\xi^u &:=& \mathring{\xi}^u(u,x^A) = -H^{-1}\partial_u \Phi , \\
\xi^A &:=& \mathring{\xi}^A(u,x^A)  +O(r^{-1}) = \mathring{\epsilon}^{AB}\partial_B \Psi+H \mathring{q}^{AB}\partial_B\Phi +O(r^{-1}), 
\end{eqnarray}    
\end{subequations}
where the subleading terms $O(r^{-1})$ in $\xi^A$ and the entire $\xi^r$ are uniquely fixed by the gauge fixing conditions. Here $\Psi=\Psi(\theta,\phi)$ is an arbitrary function over the sphere (Its $\ell=0$ harmonic can be set to zero without loss of generality). The function  $\Phi=\Phi(u,\theta,\phi)$ obeys $\partial_u^2 \Phi + \frac{H^2}{2}\mathring D_A \mathring D^A \Phi = 0$. It is a linear combination of the special solution $H u$ as well as waves in the spherical harmonic decomposition $Y_{\ell m}(\theta,\phi)e^{\pm H \omega_{\ell} u}$ with $\ell \geq 1$ and frequencies $\omega_{\ell}:=\sqrt{\frac{\ell(\ell+1)}{2}} $. The commutator between two generators $(\mathring \xi_1^u,\mathring \xi_1^A)$ and $(\mathring \xi_2^u,\mathring \xi_2^A)$ defines a generator $(\mathring \xi_3^u,\mathring \xi_3^A)$ given by 
\begin{align}
\mathring \xi^u_3 &= \mathring \xi_1^A \partial_A \mathring \xi_2^u+\frac{1}{2} \mathring \xi_1^u \mathring D_A \mathring \xi_2^A - (1 \leftrightarrow 2), \\
\mathring \xi_3^A &= \mathring \xi_1^B \partial_B \mathring \xi_2^A - H^2 \mathring \xi_1^u \mathring q^{AB} \partial_B \mathring \xi_2^u - (1 \leftrightarrow 2).  
\end{align}
This defines the $\Lambda$-BMS algebra. The 10 solutions of 
\begin{equation}
\delta_{\xi(\Phi,\Psi)} \mathring{q}_{AB} 
 := \mathring{D}_A \mathring{\xi}_B+\mathring{D}_B \mathring{\xi}_A-\mathring{q}_{AB}\mathring{D}_C \mathring{\xi}^C = 0
\end{equation}
form the $SO(1,4)$ algebra. The 10 generators are associated with $\Phi=Hu$ (time translations), $\Psi$ a combination of the 3 $\ell=1$ harmonics (rotations) and $\Phi$ a combination of the 6 $\ell=1$ waves. The other generators lead to $\delta_{\xi(\Phi,\Psi)} \mathring{q}_{AB} \neq 0$. For each harmonic $\ell$, there are $2\ell+1$ ($u$-independent) parity odd modes and $4\ell+2$ ($u$-dependent) parity even modes. There are twice as many modes in the electric sector as compared with the magnetic sector because there are two branches of solutions in the electric sector (see the $\pm H \omega_\ell u$ dependence above).  

\subsection{Bondi gauge in perturbation theory around de Sitter}

Before discussing perturbation theory around the de Sitter background, we will review the background itself. 
The de Sitter metric was written in conformally flat coordinates $x^\mu=(\eta,x^i)$ with $\rho=\sqrt{x^i x^i}$ in Eq. \eqref{background}. The coordinate transformation from conformally flat coordinates $x^\mu$ to  Bondi-Sachs coordinates $X^\mu=(u,r, x^{A})$  is 
\bea \label{21XI22.01}
\eta=- \frac{e^{-Hu}}{H(1+Hr)}, \qquad \rho=\frac{re^{-Hu}}{1+Hr},
\eea
with angles $x^A$ unchanged. The inverse transformation is 
\bea \label{ur}
u=-\frac{1}{H}{\ln\bigg(H (\rho-\eta)\bigg)}, \qquad  r=-\frac{\rho}{ H \eta}.
\eea 
The location $\rho=\eta$ (which actually corresponds to both $\rho=0=\eta$, as $\rho \geq 0, \eta \leq 0$) or, equivalently, $u=+\infty$ is excluded in Bondi gauge. The future boundary $\scri^+$ in Bondi gauge covers a $\mathbb R \times S^2$ region of $\mathbb R^3$ without the origin $\rho=\eta$, which corresponds to the upper left corner of Figure \ref{PoincarePatch}.

In Bondi-Sachs coordinates, the de Sitter metric takes the form
\bea 
\bar g_{\mu\nu} dX^\mu dX^\nu :=(H^{2}r^{2}-1)du^{2}-2du dr+r^{2} \mathring{q}_{AB}dx^{A}dx^{B}. 
\eea
The metric obeys also the boundary gauge \eqref{boundarygauge}. The non-vanishing inverse metric components are $\bar{g}^{ur}=-1$, $\bar{g}^{rr}=-H^{2}r^{2}+1$ and $  \bar{g}^{AB}=r^{-2} \mathring{q}^{AB}$. The Killing vector \eqref{time_translation} reads as $T^\mu \partial_\mu = \partial_u$.

We now would like to relate linearized perturbations in generalized harmonic gauge $h_{\mu\nu}(x^\alpha)$ with those in Bondi gauge which we will denote as $H_{\mu\nu}(X^\alpha)$. Let us denote the coordinate transformation between the two sets of coordinates as $x^\mu = \bar x^\mu(X) + \zeta^\mu(X)$ where $\bar x^\mu(X)$ is the coordinate transformation of the background \eqref{21XI22.01} and $\zeta^\mu$ is linear in the perturbation. The Bondi perturbation is given by 
\begin{align}
H_{\mu\nu}(X) &= h_{\mu\nu}(X)+\mathcal L_{\zeta}\bar g_{\mu\nu}(X),\qquad h_{\mu\nu}(X):=\frac{\partial \bar x^\alpha}{\partial X^\mu} \frac{\partial \bar x^\beta}{\partial X^\nu} h_{\alpha\beta}(\bar x(X)).\label{hbo}
\end{align}
The non-trivial part of the Jacobian matrix $\frac{\partial \bar x^\mu}{\partial X^\nu}$ is given explicitly by
\bea
\begin{bmatrix}
\frac{\partial \eta}{\partial u} & \frac{\partial \eta}{\partial r} \\
\frac{\partial \rho}{\partial u} & \frac{\partial \rho}{\partial r} 
\end{bmatrix}
=
\begin{bmatrix}
\frac{e^{-Hu}}{1+Hr} & \frac{e^{-Hu}}{(1+Hr)^{2}} \\
\frac{-Hr e^{-Hu}}{1+Hr} & \frac{e^{-Hu}}{(1+Hr)^{2}}
\end{bmatrix}.
\eea

We introduce $\perp^{ij}:= \delta^{ij} -n^{i}n^{j}$ and $e^i_A = \frac{\partial n^i}{\partial x^A}$ which obeys $e^i_A n_i=0$. We also define the symmetric tracefree tensor  $e^i_{\langle A}e^j_{B \rangle} :=e^i_{(A} e^j_{B)} -\frac{1}{2} \mathring{q}_{AB} \perp^{ij}$. The metric perturbation $h_{\mu\nu}(X)$ defined in Eq. \eqref{hbo} reads as 
\bea \nonumber 
h_{uu}(X)
&=& \bigg(\frac{\partial \eta}{\partial u}\bigg)^{2}h_{00} +2 n^i \frac{\partial \eta}{\partial u} \frac{\partial \rho}{\partial u}h_{0i}+ n^i n^j \bigg(\frac{\partial \rho}{\partial u}\bigg)^{2} h_{ij}\\
&=& \frac{1}{2} \bigg(1+H^2 r^2\bigg) \hat \chi - 2 H r n^{i} \chi_{0i} - H^{2} r^{2} \perp^{ij} \chi_{ij}, \\ \nonumber 
h_{ur}(X)
&=& \frac{\partial \eta}{\partial u}\frac{\partial \eta}{\partial r} h_{00}+n^i\bigg(\frac{\partial \eta}{\partial u} \frac{\partial \rho}{\partial r}+ \frac{\partial \eta}{\partial r} \frac{\partial \rho}{\partial u}\bigg)h_{0i}+n^{i} n^j \frac{\partial \rho }{\partial u}\frac{\partial \rho}{\partial r} h_{ij}\\
&=& \frac{1}{2}\bigg(\frac{1-Hr}{1+Hr}\bigg) \bigg({\hat {\chi}}+ 2n^{i} \chi_{0i}\bigg) + \bigg( \frac{Hr}{1+Hr}\bigg) \perp^{ij} \chi_{ij},\\
h_{uA}(X)
&=& \rho \frac{\partial \eta}{\partial u} \frac{\partial n^i}{\partial x^A} h_{0i}+\rho n^i \frac{\partial \rho}{\partial u} \frac{\partial n^j}{\partial x^A} h_{ij}=r e^i_A \chi_{0i} - Hr^{2} n^{i}e^j_A\chi_{ij},\\ \nonumber 
h_{rr}(X)
&=& \bigg(\frac{\partial \eta}{\partial r}\bigg)^{2} h_{00}+2 n^{i} \frac{\partial \eta}{\partial r} \frac{\partial \rho}{\partial r} h_{0i}+n^{i}n^{j} \bigg(\frac{\partial \rho}{\partial r}\bigg)^{2} h_{ij} \\
&=& \frac{1}{(1+Hr)^{2}}\bigg[\hat{\chi}+2n^{i} \chi_{0i}-\perp^{ij} \chi_{ij}\bigg],\\
h_{rA}(X)
&=& \rho \frac{\partial \eta}{\partial r} \frac{\partial n^i}{\partial x^A} h_{0i}+\rho n^i \frac{\partial \rho}{\partial r} \frac{\partial n^j}{\partial x^A} h_{ij}=\frac{r}{1+H r} e^i_A (\chi_{0i}+n_j \chi_{ij}),\\
h_{AB}(X)
&=& \rho^2 e^i_{A} e^j_{B} h_{ij}= r^2 (e^i_{(A} e^j_{B)} - \mathring{q}_{AB}\delta_{ij})\chi_{ij}+\frac{1}{2}r^2\mathring{q}_{AB}\hat\chi. 
\eea

Linear perturbations around de Sitter in Bondi-Sachs gauge obey 
\bea 
H_{rr}=0,\qquad H_{rA}=0,\qquad \partial_r( \mathring{q}^{AB} H_{AB})=0. \label{Hg}
\eea
The supplementary boundary gauge fixing conditions following from Eq. \eqref{boundarygauge} are 
\begin{align}
\text{lim}_{r \rightarrow \infty} (r^{-2}\mathring{q}^{AB}H_{AB} )=0, \label{boundarygauge2}
\end{align}
and
\begin{align}
\text{lim}_{r \rightarrow \infty} (r^{-2}H_{uu})=0,\quad \text{lim}_{r \rightarrow \infty} (r^{-2}H_{uA} )=0. \label{boundarygauge3}
\end{align}
These two conditions are equivalent to the first two conditions of Eq. \eqref{boundarygauge} in the notation of \cite{Compere:2019bua}. Einstein's equations then imply that
\begin{equation}\label{Hur}
H_{ur}= h_{ur}+(H^2 r^2-1)\partial_r \zeta^u-\partial_u \zeta^u - \partial_r \zeta^r =0,
\end{equation}
as will be checked below. In other words, at linear order, imposing Bondi gauge and boundary gauge conditions \eqref{boundarygauge} implies that the linearized metric is also in Newman-Unti gauge, as demonstrated in \cite{Compere:2019bua} (this is also true in the flat limit, see the explicit linear Bondi metric in \cite{Blanchet:2020ngx}).  The solution to the gauge fixing conditions \eqref{Hg} and \eqref{boundarygauge2}  is 
\begin{eqnarray}
\zeta^{u} (u,r,x^{A})&=& \xi^{u}(u,x^{A}) +\frac{1}{2} \int^{r} dr' h_{rr}(u,r',x^{A}), \\ 
\zeta^{A}(u,r,x^{A}) &=& \xi^{A}(u,x^{A}) +\mathring{q}^{AB} \int^{r} \frac{dr'}{r^{\prime 2}} (\partial_{B} \zeta^{u}(u,r',x^{A}) -h_{rB}(u,r',x^{A})), \\
\zeta^{r}(u,r,x^{A}) &=& -\frac{r}{2} \mathring{D}_{B} \zeta^{B} -\frac{1}{4r} \mathring{q}^{AB} h_{AB}(u,r,x^{A}). 
\end{eqnarray}
The non-vanishing Bondi metric components are then given by 
\begin{eqnarray}\label{Bm1}
H_{uu} &=& h_{uu}+2H^2 r \zeta^r-2 \partial_u \zeta^r+2(H^2 r^2-1)\partial_u \zeta^u,\\
H_{uA} &=& h_{uA}+r^2 \mathring{q}_{AB}\partial_u \zeta^B-\partial_A \zeta^r + (H^2 r^2-1)\partial_A \zeta^u , \\
H_{AB} &=& h_{AB}+r^2 (\mathring{q}_{AC}\mathring{D}_{B} \zeta^{C}+\mathring{q}_{BC}\mathring{D}_{A} \zeta^C-\mathring{q}_{AB} \mathring{D}_{C} \zeta^{C})- \frac{1}{2} \mathring{q}_{AB}\mathring{q}^{CD}h_{CD} . \label{Bm4}
\end{eqnarray}
Since we need to know the asymptotic behavior of $H_{uu}$ and $H_{uA}$ to impose the remaining gauge fixing conditions \eqref{boundarygauge3}, we will only impose them for the specific case of the quadrupolar metric in the next section\footnote{If one imposes $\text{lim}_{r \rightarrow \infty}(r^{-2}H_{AB})=0$ instead of $\text{lim}_{r \rightarrow \infty}(r^{-2}H_{uA})=0$ while keeping $\text{lim}_{r \rightarrow \infty}(r^{-2}H_{uu})=0$, one deduces from Eqs. \eqref{Bm1}-\eqref{Bm4} that the residual coordinate transformations are generated by one arbitrary function of the angles $x^A$ in $\xi^u(u,x^A)$ and 6 arbitrary functions of $u$ respectively multiplying the conformal Killing vectors $\mathring \zeta^A(x^A)$ of $\mathring q_{AB}$. The asymptotic symmetry algebra $\mathbb R \times SO(3)$ claimed in \cite{Bonga:2023eml} is therefore part of an infinite-dimensional algebra which is a semi-direct product of $\text{vect}(S^2)$ with an affine extension of $SO(3,1)$. This remains to be investigated in detail. We will not consider such alternative boundary conditions further.}.

\subsection{Quadrupolar Bondi metric and flat limit}

The fields $\hat \chi$, $\chi_{0i}$ and $\chi_{ij}$ as defined in Eqs. \eqref{finalchi1}, \eqref{finalchi0i} and \eqref{finalchij} are functional of $\eta_{\text{ret}}=-H^{-1}e^{-H u}$ or, equivalently, $u$ and $\rho a(\eta_\text{ret})=\frac{r}{1+H r}$ or, equivalently, $r$. They are also functional of the multipole moments that only depend upon the retarded conformal time, or equivalently, of $u$\footnote{In case Lie derivatives are being used, note that we can also substitute
$\mathcal L_T Q_{ij} (\eta - \rho) = -H(\eta \partial_{\eta}+\rho \partial_{\rho})Q_{ij}(\eta-\rho)-2HQ_{ij} = \partial_{u} Q_{ij}-2HQ_{ij}.$}. 

After a straightforward but lengthy computation, we obtain the explicit Bondi metric components \eqref{Bm1}-\eqref{Bm4}. The boundary gauge fixing equations \eqref{boundarygauge3} become 
\begin{eqnarray}\label{eqbnd1}
\partial_u \xi^u-\frac{1}{2}\mathring{D}_A \xi^A &=&\frac{1}{4}(\delta^{ij}-3n^i n^j)\chi_{ij}^{(0)},\\ \label{eqbnd2}
\partial_u \xi^A+H^2 \mathring{q}^{AB}\partial_B \xi^u &=&H n^i e^j_A \chi_{ij}^{(0)}. %
\end{eqnarray}
where $\chi^{(0)}_{ij}$ was defined in Eq. \eqref{chi0}. The boundary metric defined as $q_{AB} := \text{lim}_{r \rightarrow \infty}(r^{-2}(\bar g_{AB}+H_{AB}))$ is given by 
\begin{equation}
q_{AB}= \mathring q_{AB} + \chi^{(0)}_{ij}e^i_{\langle A} e^j_{B \rangle}+2\mathring q_{C \langle A}D_{B \rangle} \xi^C. 
\end{equation}

The homogeneous solutions to the boundary gauge fixing equations \eqref{eqbnd1}-\eqref{eqbnd2} are the $\Lambda$-BMS asymptotic symmetries described in Section \ref{sec:BMS}. Let us call its generators $\mathring{\xi}^u(u,x^A)$ and $\mathring{\xi}^A(u,x^A)$ which obey 
\begin{equation} \label{eq:hom-xi}
\partial_u \mathring{\xi}^u-\frac{1}{2}\mathring{D}_A \mathring{\xi}^A =0,\qquad \partial_u \mathring{\xi}^A+H^2 \mathring{q}^{AB}\partial_B \mathring{\xi}^u=0.
\end{equation}
At quadrupolar order, the only angular dependency in the right-hand side of \eqref{eqbnd1}-\eqref{eqbnd2} is contained in $n^i$ and $e^i_A$. The solution to the boundary gauge fixing conditions takes the form 
\begin{eqnarray} \label{Bdgauge1}
\xi^u &\!\!=\!\!& \mathring{\xi}^u(u,x^A)+n_i \zeta_i+ (\delta^{ij}-3n^i n^j)\left(\frac{1}{4}\zeta_{ij}+\frac{H^2}{2}Q^{(\rho)}_{ij}-\frac{\chi_{ij}(-\infty)}{6H}\right)+\frac{H^2}{2}n^i n^j Q^{(p)}_{ij},\\ \label{Bdgauge2}
\xi^A &\!\!=\!\!& \mathring{\xi}^A(u,x^A)+e^A_i (-\partial_u \zeta_i+\frac{H^2}{2}Q^{(p)}_i)+ \frac{1}{2}\mathring{q}^{AB}e_B^i n^j \left( \partial_u \zeta_{ij}+2H^2\partial_u Q^{(\rho)}_{ij}-\chi_{ij}^{(0)}(u,x^A)\right)\nonumber\\
&&+\frac{H^2}{2}\epsilon_{ijk}e^A_i n^{j}n^{l}(K_{kl}+2 H \int^u du'K_{kl}(u')), 
\end{eqnarray}
where $\zeta_{i}(u)$ and $\zeta_{ij}(u)$ obey the following wave equations with sources
\begin{align}\label{inhom}
\partial_u^2 \zeta_{i} - H^2 \zeta_{i} &=H^2  (\partial_u +H)Q^{(p)}_i , \\
\partial_u^2 \zeta_{ij} -3 H^2 \zeta_{ij} &=-2H^4 Q^{(\rho+p)}_{ij}.
\end{align}
The particular solutions $\zeta_{i}(u)$, $\zeta_{ij}(u)$ can be written explicitly using the method of variations of constants. The boundary gauge fixing conditions are then solved. 

We find that the final Bondi metric component $H_{ur}=\frac{1}{2} \mathring{D}_{A} \mathring\xi^{A} -\partial_{u} \mathring\xi^{u}=0$ and Eq. \eqref{Hur} is obeyed, which is a non-trivial check of our expressions. 

The final Bondi metric component $\bar g_{AB}+H_{AB}$ at quadrupolar order takes the form
\begin{equation}
\bar g_{AB}+ H_{AB} = r^2 q_{AB}+ r C_{AB} + \frac{1}{r} E_{AB},
\end{equation}
where the boundary metric $q_{AB}$, the shear $C_{AB}$ and the tensor $E_{AB}$ are given by
\begin{align}
q_{AB} &= \mathring{q}_{AB}+ 2 \mathring q_{C \langle A}\mathring{D}_{B \rangle} \mathring{\xi}^C + e^i_{\langle A} e^j_{B \rangle} \bigg(\partial_u \zeta_{ij}+2H^2 \partial_u Q^{(\rho+p)}_{ij}\nonumber\\
&\qquad\qquad \qquad\qquad\qquad\qquad\qquad+2H^2 \epsilon_{i kl}n_k(K_{jl}+H\int^u du'K_{jl}(u'))\bigg), \label{qABBondi}\\
C_{AB} &= -2\mathring{D}_{\langle A} \mathring{D}_{B\rangle}\mathring \xi^{u}  +e_{\langle A}^i e_{B \rangle}^j \left( 3\zeta_{ij} +2 (\partial_u^2 -H^2)Q^{(\rho+p)}_{ij}+2\epsilon_{ikl}n_k (\partial_u+H)K_{lj}\right),\\
E_{AB} & = 2 e_{\langle A}^i e_{B \rangle}^j \left( Q^{(\rho+p)}_{ij} + \epsilon_{ikl}n_k J_{jl}\right). 
\end{align}
The $r^0$ term vanishes as expected from Einstein's equations, which is again a non-trivial check of our computation. Also, we have $\partial_{u}q_{AB}=H^{2} C_{AB}$ which is consistent with Einstein's equations in our gauge.

The component $g_{uu} = \bar g_{uu}+H_{uu}$ reads as
\begin{align}
g_{uu} = \bar g_{uu}+H_{uu}&=H^2 r^2+(\delta^{ij}-3n^i n^j)\partial_u (\zeta_{ij}+2H^2 Q^{(\rho+p)}_{ij})-(\mathring D^C \mathring D_C +2)\partial_u \mathring\xi^u-1\nonumber\\ 
&+\frac{2M}{r}+\frac{2N}{r^2}+\frac{(3n^{i}n^{j}-\delta^{ij})Q^{(\rho+p)}_{ij}}{r^3},
\end{align}
where 
\begin{align}
M &= Q^{(\rho)}-2 Q^{(p)} -3n^{i}(P_{i}-HQ_{i}^{(\rho)}-H^2 P_{i \vert kk})+3 n^{i}n^{j}(2 S_{ij}-H P_{i|j})\nonumber\\
& + (\delta^{ij}-3n^{i} n^{j}) (3 H P_{i \vert j}-3H^2 Q^{(\rho)}_{ij} +3 H^2 Q^{(p)}_{ij}+H \partial_u Q^{(p)}_{ij}-\partial_u^2 Q^{(p)}_{ij}),\\
N &=  n_i (Q^{(\rho)}_i+H P_{i \vert kk})-(\delta_{ij}-3 n_{i}n_{j}) \partial_{u} Q_{ij}^{(\rho +p)}.
\end{align}
Finally $H_{uA}$ is given by 
\begin{align}
H_{uA} &= -e_A^i n^j (H^{-2}\partial_u^2 \zeta_{ij}+2\partial_u^2 Q^{(\rho+p)}_{ij}+2 \epsilon_{i kl}n_k(\partial_u+H) K_{jl})-\frac{1}{2}\mathring D_A (\mathring D_C \mathring D^C +2)\mathring\xi^u \nonumber\\
&+ \frac{2 e_A^i N_i }{r}+\frac{ 3e_A^i n^j ( Q_{ij}^{(\rho+p)}+ \epsilon_{ikl}n_k J_{lj})}{r^2}, 
\end{align}
where
\begin{align}
N_i &= Q_i^{(\rho)}+H P_{i \vert kk} +n^j (\epsilon_{ijk}J_k+2\partial_u Q^{(\rho+p)}_{ij})-2 \epsilon_{ijk}n_j n_l (K_{kl}- H J_{kl}).
\end{align}

Quite remarkably, all metric components only contain a finite number of terms in the $1/r$ expansion, which is directly related to the quadrupolar truncation, and do not contain any $\log r$ terms. These two features also arise in the flat case \cite{Blanchet:2020ngx}.

The flat limit is taken as $H \mapsto 0$, which implies $\chi_{ij}(-\infty) \mapsto 0$ and $\zeta_{ij} \mapsto 0$. Note that $H^{-2}\partial_u^2 \zeta_{ij} \mapsto 4 S_{ij}$, $S_{ij} \mapsto \frac{1}{2}\ddot Q_{ij}^{(\rho)}$, $P_{(i \vert j)} \mapsto -\frac{1}{2}\dot Q^{(\rho)}_{ij}$, $P_{[i \vert j]} \mapsto \frac{1}{2}\epsilon_{ijk}J_k$, $P_i \mapsto - \dot{Q}^{(\rho)}_i$, $K_{ij} \mapsto - \dot J_{ij}$. As a consequence of Eq. \eqref{conseq2} we also have $\ddot Q^{(\rho)}_{kk} \mapsto 2 Q^{(p)}$. In the flat limit, the $\Lambda$-BMS group reduces to the generalized BMS group and we have $\mathring \xi^A=Y^A(x^C)$, $\mathring \xi^u=T(x^A)+\frac{u}{2}\mathring D_AY^A$ in terms of the three arbitray functions over the sphere $T$ associated with supertranslations and $Y^A$ associated with super-Lorentz transformations. In the flat limit, the Bondi metric up to quadrupolar order is given by 
\begin{align} \nonumber
g_{uu} & = -1-\frac{1}{2}(\mathring{D}_{A} \mathring{D}^{A}+2)\mathring{D}_{B}Y^B +\frac{2}{r}(Q^{(\rho)}+3n^{i}\dot Q_{i}^{(\rho)} +(3n^i n^j -\delta^{ij}) \ddot Q^{(\rho + p)}_{ij})\\  
&+\frac{2}{r^2} (n^i Q^{(\rho)}_{i}-(\delta^{ij}-3n^i n^j)\dot Q^{(\rho +p)}_{ij}) +\frac{1}{r^3} (3n^{i}n^{j}-\delta^{ij}) Q_{ij}^{(\rho +p)},\\
g_{uA} &=  -\frac{1}{2}\mathring{D}_{A}(\mathring{D}_{B}\mathring{D}^{B}+2)\mathring\xi^{u} -2 e^{i}_{A} n^{j}(\ddot Q_{ij}^{(\rho + p)}-  \epsilon_{i k l }n_k\ddot J_{jl})\nonumber\\
&+ \frac{2 e_A^i(Q^{(\rho)}_i+n_j\epsilon_{ijk}(J_k+2n_l \dot J_{kl})+2n_j\dot Q^{(\rho+p)}_{ij})}{r}+\frac{ 3e_A^i n^j (Q_{ij}^{(\rho+p)}+ \epsilon_{ikl}n_k J_{lj})}{r^2} , \\
g_{AB} &= r^{2} \bigg(\mathring q_{AB}+2 \mathring{D}_{\langle A} \mathring \xi_{B\rangle} -\frac{2}{r} \mathring{D}_{\langle A} \mathring{D}_{B\rangle}\mathring \xi^{u}+\frac{2}{r} e^{i}_{\langle A}e^{j}_{ B \rangle} (\ddot Q_{ij}^{(\rho + p)}- \epsilon_{ikl}n_k \ddot J_{lj}) \nonumber\\
&+\frac{2}{r^{3}}e^{i}_{\langle A}e^{j}_{ B \rangle} ( Q_{ij}^{(\rho + p)}+\epsilon_{ikl}n_k J_{jl})\bigg). 
\end{align}
The results then agree with Eq. (23) of \cite{Blanchet:2020ngx} when restricted up to quadrupolar order after identifying the STF canonical multipole moments $M_L$, $S_L$ for $\ell=0,1,2$ as
\begin{align}
M_\emptyset &= Q^{(\rho)}, \qquad 
M_i  = Q^{(\rho)}_i, \qquad 
M_{ij} =  Q^{(\rho + p)}_{ij}-\frac{1}{3}\delta_{ij}  Q^{(\rho + p)}_{kk}, \\
S_i &= J_i , \qquad 
S_{ij} = \frac{3}{4}J_{ij}. 
\end{align}

\section{Linear cosmological displacement memory}
\label{sec:memory}

\subsection{Description in harmonic gauge}
\label{sec:memory1}

We consider the situation depicted in Figure \ref{PoincarePatch} (see also Figure \ref{dSMemory}). In the limit to $\scri^+$ (reached as $\eta \rightarrow 0$ at fixed $\rho$), one has $a(\eta_{\text{ret}})^{-1}\rightarrow H \rho$ and it is immediate to see that the scalar and vector perturbations given in Eqs. \eqref{finalchi1} and \eqref{finalchi0i} vanish $\hat\chi(\eta, x) =0$, $\chi_{0i}(\eta, x) =0$. Now, the late time field  $\chi_{ij}$ given in Eq. \eqref{finalchij} is non-vanishing: 
\bea  \label{16IX22.01}
\chi_{ij}(\eta=0,\rho)=\chi_{ij}^{(0)}(-\rho,x^A),
\eea
where $\chi_{ij}^{(0)}(-\rho,x^A)$ is defined in Eq. \eqref{chi0} with $\eta_\text{ret}$ substituted with $-\rho$. The late-time imprint of radiation in asymptotically de Sitter spacetimes at leading order in the metric expansion is a feature of gravity with a positive cosmological constant. 

The source is located at $\rho = 0$ at all times $\eta$. Let us first discuss the case where the source is stationary at early and late times, 
\begin{equation}
\partial_t (\text{any multipole})=0,\qquad \text{as } \eta > \eta_{\text{f}}\; \text{ or }\; \eta < \eta_{\text{i}}, \label{statass}
\end{equation}
where $\eta_{\text{i}}$ is an early time cutoff and $\eta_{\text{f}}$ is a late time cutoff. All fields that propagate along the lightcone will then only vary over a fixed interval on $\mathcal I^+$. More precisely, the $\partial_t$ derivative of all instantaneous terms in $\chi_{ij}^{(0)}(\rho,x^A)$ will vanish outside of the $\rho$ interval $[-\eta_{\text{f}},-\eta_\text{i}]$. The stationarity assumption \eqref{statass} also implies that there is no radiation originating from the cosmological horizon at $\eta=-\infty$. The expression for $\chi_{ij}(-\infty)$ \eqref{chiijinf}, the only tail effect at linear order (that propagates inside the lightcone), becomes 
\begin{equation}\label{tailval}
\chi_{ij}(-\infty)=2H^3 (2 Q^{(\rho)}_{ij}(\eta_{\text{i}})-Q^{(p)}_{ij}(\eta_{\text{i}})). 
\end{equation}
In both non-radiative regions $\rho < -\eta_{\text{f}}$ and $\rho > -\eta_\text{i}$ of $\mathcal I^+$ the expression for $\chi_{ij}(\eta=0)$ \eqref{chi0eq} for such $\rho=\rho_{\text{non-rad}}$ (where $\rho_{\text{non-rad}}$ is a constant equal to either $-\eta_\text{i}$ or $-\eta_\text{f}$) is 
\begin{equation}
\chi_{ij}(\eta=0,\rho_{\text{non-rad}})= \chi_{ij}(-\infty)+2H^2(-2H Q^{(\rho)}_{ij}+n_k \epsilon_{kl(i}K_{j)l}-n_{(i}Q^{(p)}_{j)}+\delta_{ij}n_k Q^{(p)}_k)\vert_{\eta=-\rho_{\text{non-rad}}},
\end{equation}
where all moments are evaluated at $\eta=-\rho_{\text{non-rad}}$. The $TT$ transverse-traceless part of $\chi_{i j}(\eta=0,\rho_{\text{non-rad}})$ defined as the part of $\chi_{i j}$ which is traceless and divergence-free with respect to $\partial_i$  (that obeys $\frac{\partial}{\partial n_i}\chi^{TT}_{i j}=0$ given the absence of $\rho$ dependence) is given by 
\begin{equation}\label{chiTTnonrad}
\chi_{ij}^{TT}(\eta=0,\rho_{\text{non-rad}})= \chi_{\langle ij \rangle}(-\infty)+4H^3(- Q^{(\rho)}_{\langle ij \rangle}+\frac{1}{2}n_k \epsilon_{kl(i}J_{j)l})\vert_{\eta=-\rho_{\text{non-rad}}}, 
\end{equation}
where we used Eq. \eqref{consJK}. It contains both the mass and the current quadrupole. 

Hence, if the source is only active at $\rho=0$ between some initial retarded time $\eta_{\text{i}}$ and final time $\eta_{\text{f}}$, the difference between the final and initial fields at the corresponding $\rho=-\eta_{\text{i}}$ and $\rho=-\eta_{\text{f}}$ at $\scri^+$ is
\begin{align}
\Delta\chi^{TT}_{ij} (x^A) &:= \chi^{TT}_{ij}[\eta=0,-\eta_{\text{i}}]-\chi^{TT}_{ij}[\eta=0,-\eta_{\text{f}}] 
\\
&= 4H^{3}\bigg[-Q_{\langle ij \rangle}^{(\rho)} |_{\eta=\eta_{\text{f}}}+\frac{1}{2}n_k \epsilon_{kl(i}J_{j)l}|_{\eta=\eta_{\text{f}}} + Q_{\langle ij \rangle}^{(\rho)} |_{\eta=\eta_{\text{i}}}-\frac{1}{2}n_k \epsilon_{kl(i}J_{j)l}|_{\eta=\eta_{\text{i}}}\bigg].\label{memory}
\end{align}
If the source is stationary at all times, $\Delta\chi^{TT}_{ij}$ vanishes. The tail term \eqref{tailval} being a constant, it cancels out in the difference. While $\chi_{ij}^{TT}(\eta=0,\rho_{\text{non-rad}})$ is gauge dependent in each non-radiative region, see Eq. \eqref{22XI22.01}, the difference between the two non-radiative regions $\Delta\chi^{TT}_{ij}$ is gauge invariant. This is the cosmological displacement memory effect, a permanent change of the leading boundary metric of asymptotically de Sitter spacetimes due to radiation reaching $\scri^+$.

Contrary to the asymptotically flat displacement memory effect, the cosmological memory effect leads to an order one effect at late time which does not decrease with the radius. This effect was noticed by Chu \cite{Chu:2015yua,Chu:2016qxp}. 
Note however that Chu argues that the $TT$ part of $\chi_{ij}(\eta =0,\rho)$ is gauge invariant, which is not the case. In vacuum regions where there is no matter, residual transformations transform $\chi_{ij}^{TT}$ as Eq. \eqref{22XI22.01} as proven in Section \ref{sec:gauge}. In particular, the parameters $\Xi^i=b_{ij}x_j$ with $b_{ij}$ constant, symmetric and traceless trivially obey the equation $\Delta \Xi^i=-\frac{1}{3}\partial_i \partial_j \Xi^j$ and shift $\chi_{ij}^{TT}$ by a general constant term 
\begin{equation}
\delta_{\xi} \chi_{ij}^{TT}=  b_{ij}. 
\end{equation}
As a consequence, either the initial or the final constant values of $\chi_{ij}^{TT}$ can be set to zero, but not both. As usual in memory effects, the difference between the initial and final stationary states \eqref{memory} is the gauge invariant memory. 

The computations of Chu \cite{Chu:2015yua,Chu:2016qxp} only considered the mass quadrupolar radiation. Our expression \eqref{memory} also includes the current quadrupolar radiation which brings one further qualitative feature: the memory is angle-dependent. Moreover, contrary to the parity-even sector, the odd-parity sector cannot be gauged away in non-radiative regions because there is no parameter $\Xi^i$ in \eqref{solgensym} such that $\delta_{\Xi}\chi_{ij}^{TT}$ cancels the last term in Eq. \eqref{chiTTnonrad}. There is therefore no correspondence between the odd-parity displacement memory and asymptotic symmetries in generalized harmonic gauge.

Let us now relax the stationarity assumption \eqref{statass}. If the source is again only active at $\rho=0$ between some initial retarded time $\eta_{\text{i}}$ and final time $\eta_{\text{f}}$, the difference between the final and initial fields at the corresponding $\rho=-\eta_{\text{i}}$ and $\rho=-\eta_{\text{f}}$ at $\scri^+$ is
\begin{align}
\Delta\chi^{TT}_{ij} (x^A) &:= \chi^{(0) TT}_{ij}[\eta=0,-\eta_{\text{i}}]-\chi^{(0) TT}_{ij}[\eta=0,-\eta_{\text{f}}] 
\end{align}
where $\chi^{(0)}_{ij}$ is defined in Eq. \eqref{chi0}. Let us denote $\omega_s$ the typical frequency of the source. If one assumes stationarity in the past, as it is performed in the PN/PM formalism in the flat case, we see from Eq.  \eqref{chi0}
 that the boundary metric will contain terms of order $H^2 \omega_s$. If one further admits non-stationary terms in the past, the boundary metric will admit terms linear in $H \omega_s^2$ arising from $\chi_{ij}(-\infty)$, see Eq.  \eqref{chiijinf}.  In the general non-stationary case, the boundary metric will contain terms of order $H^3$, $H^2 \omega_s$ and $H \omega_s^2$, where the last term will dominate the effect for localized sources which admit $\omega_s \gg  H$.


\subsection{Description in $\Lambda$-BMS gauge}

\begin{figure}[htb]
\centering
\begin{center}
\includegraphics[scale=0.35]{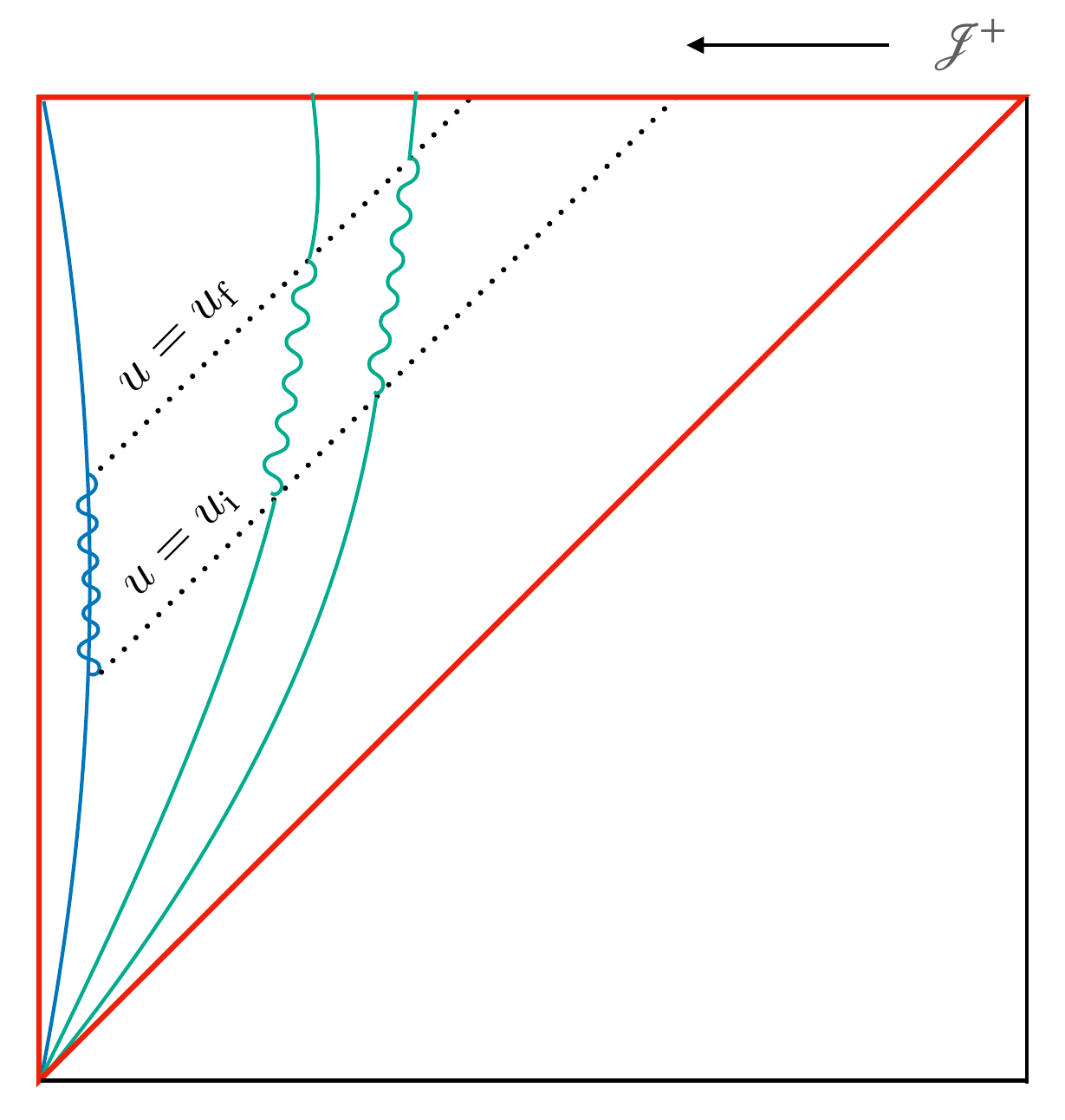} 
\caption{Bondi-Sachs coordinates are adapted to the future Poincaré patch of de Sitter space-time. Blue line indicates the trajectory of a spatially compact source in de Sitter background. The source is emitting gravitational radiation in the interval $(u_\text{i},u_\text{f})$. Green lines denote the trajectory of two inertial detectors. Gravitational radiation in the interval $(u_\text{i}, u_\text{f})$ causes a temporary oscillation in their relative positions followed by a permanent displacement which is a displacement memory effect.
\label{dSMemory}
}
\end{center}
\end{figure}

\begin{figure}[htb]
    \centering
    ~~~~~
        \includegraphics[width=0.75\textwidth]{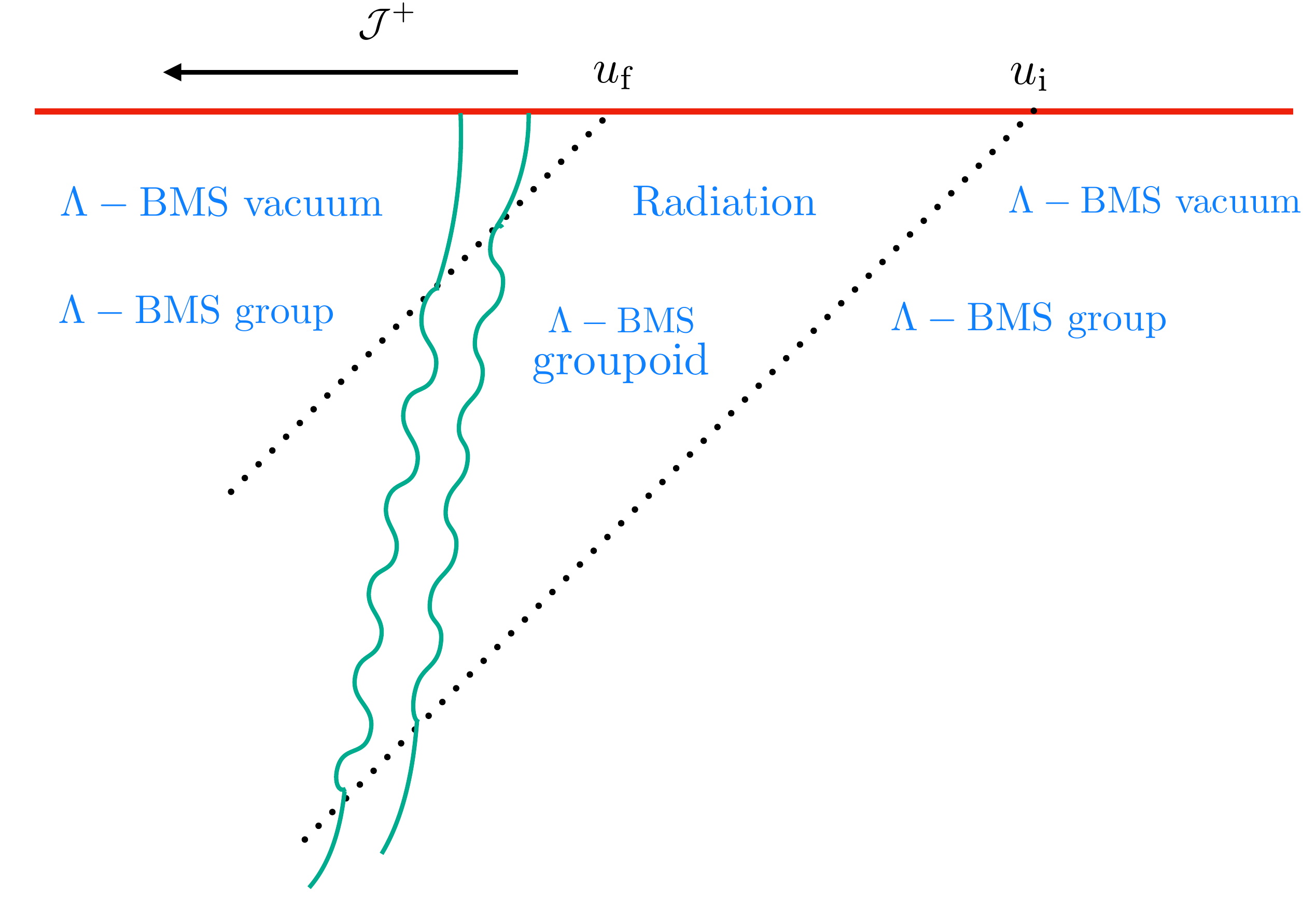}
    \qquad\qquad 
    \caption{In the region $u \in [u_\text{i},u_\text{f}]$ where radiation hits future infinity $\scri^+$, the asymptotic symmetry group is the $\Lambda$-BMS groupoid. In the regions $u < u_\text{i}$ and $u> u_\text{f}$ without radiation and in the absence of odd parity quadrupole, $\scri^+$ admits the ordinary $\Lambda$-BMS group as asymptotic symmetry group. The two vacua on either side of the radiation region are related by a $\Lambda$-BMS group element \eqref{BMSel}-\eqref{genBMS}. The even parity part of the displacement memory corresponds to such as a $\Lambda$-BMS transition.}\label{Transition}
\end{figure}

In $\Lambda$-BMS gauge \eqref{boundarygauge2}-\eqref{boundarygauge3}, the leading order metric on the celestial sphere $q_{AB}$ varies according to the radiation sent to $\scri^+$. The crucial formula for our purposes \eqref{qABBondi} gives the metric perturbation of the boundary metric
\begin{align}
\delta q_{AB} &:=  q_{AB}-\mathring{q}_{AB} = 2 \mathring{D}_{\langle A} \mathring{\xi}_{B \rangle} + e^i_{\langle A} e^j_{B \rangle} \bigg(\partial_u \zeta_{ij}+2H^2 \partial_u Q^{(\rho+p)}_{ij}\nonumber\\
&\hspace{6.5cm}+2H^2 \epsilon_{i kl}n_k(K_{jl}+H\int^u du'K_{jl}(u'))\bigg). \label{finalqAB2}
\end{align}

Let us consider again a system where the radiation reaches $\scri^+$ only in a finite interval $u \in [u_\text{i},u_\text{f}]$, which corresponds to a source emitting gravitational waves during a finite proper time interval, see Figure \ref{dSMemory}. For both $u < u_\text{i}$ and $ u > u_\text{f}$, the fields $Q^{(\rho+p)}_{ij}(u)$, $K_{ij}(u)$ are constant. We shift $u$ so that $u=0$ belong to the interval $[u_\text{i},u_\text{f}]$. In order to compute the displacement memory, we assume for simplicity that the field $Q^{(\rho+p)}_{ij}(u)$ jumps at $u=0$ as   $Q^{(\rho+p)}_{ij}(u)= Q^{(\rho+p)}_{ij}(u_i)+(Q^{(\rho+p)}_{ij}(u_f)-Q^{(\rho+p)}_{ij}(u_i))\Theta(u)$ where $\Theta(u)$ is the step function, and similarly for $K_{ij}(u)$. We can then solve the inhomogeneous wave equation \eqref{inhom} exactly as 
\begin{equation}
\zeta_{ij}(u) =\frac{2H^2}{3} \left(Q^{(\rho+p)}_{ij}(u_\text{i})+(Q^{(\rho+p)}_{ij}(u_\text{f})-Q^{(\rho+p)}_{ij}(u_\text{i}))(\cosh{(\sqrt{3}H u)}-1)\Theta(u)\right),
\end{equation}
which implies 
\begin{equation}
\partial_u\zeta_{ij}(u) =\frac{2H^3}{\sqrt{3}}(Q^{(\rho+p)}_{ij}(u_\text{f})-Q^{(\rho+p)}_{ij}(u_\text{i}))\sinh(\sqrt{3}H u)\Theta(u) . 
\end{equation}
In Eq. \eqref{finalqAB2} we consider as asymptotic symmetry vectors $\mathring \xi^A$ the two homogeneous solutions obeying \eqref{Eqsxi} that are proportional to $e^{\pm \sqrt{3}H u}$. They are obtained from the $2 \times 5$ generators 
\begin{eqnarray}
\mathring \xi^A = 2\sqrt{3}H e^A_i n_j e^{\pm \sqrt{3}H u} c_{ij},\qquad \mathring \xi^u= \pm (\delta_{ij}-3 n_i n_j)c_{ij}e^{\pm \sqrt{3}H u} \label{genBMS}
\end{eqnarray}
of the $\Lambda$-BMS group that shift $q_{AB}$ according to \eqref{finalqAB2}. Here $c_{ij}$ is symmetric and determines 5 constants. We can use the 10 homogeneous solutions to remove the $Q^{(\rho+p)}_{ij}$ dependency in $\delta q_{AB}$ at either $u < u_\text{i}$ or $u > u_\text{f}$ but not both. There is no residual gauge transformation that allows to remove the dependency in the current quadrupole moment $K_{ij}$ at either $u < u_\text{i}$ or $u > u_\text{f}$.

 The difference $\delta q_{AB}(u_\text{f})-\delta q_{AB}(u_\text{i})$ is gauge invariant and leads to the finite displacement memory effect caused by gravitational radiation reaching $\scri^+$. The even parity sector can be interpreted as a $\Lambda$-BMS transition between two vacuum states at $u=u_\text{i}$ and $u=u_\text{f}$ generated by the asymptotic symmetry group element \eqref{genBMS}. In that sense, we found the exact analogue for $\Lambda >0$ of the relationship between displacement memory effect and asymptotic symmetry group first described for $\Lambda=0$ \cite{Strominger:2014pwa}. We depict this effect in Figure \ref{Transition}. The odd parity sector cannot be interpreted as a transition between vacua in $\Lambda$-BMS gauge due to the absence of suitable symmetries. This parallels the flat odd-parity displacement memory effect \cite{Satishchandran:2019pyc}, which occurs however at one comparatively subleading order with respect to the leading order odd parity memory effect that we are describing (which becomes trivial in the flat limit). 

Let us now describe the displacement memory effect itself in more detail. We consider a generic asymptotically de Sitter metric $g_{\mu\nu}$ in Bondi gauge (see \cite{Compere:2019bua}). Timelike geodesics with $\Lambda \neq 0$ admit  $v^r = H r + O(r^0)$. We consider a timelike geodesic of tangent vector
\begin{align}
v^\mu   = \left(v_{-1}^u+\frac{v_0^u}{r}+\frac{v_1^u}{r^2}+O(r^{-3}),H r +v_1^r +O(r^{-1}),\frac{v_0^A}{r}+\frac{v_1^A}{r^2}+\frac{v_2^A}{r^3} +O(r^{-4}) \right).
\end{align}
The normalization condition $v^{\mu}v_{\mu}=-1$ implies $\Lambda v_{-1}^u =0$, which is solved in the case $\Lambda \neq 0$ as $v_{-1}^u = 0$. The geodesic equation then implies $v_0^A=0$ and $v_0^u = \pm H^{-1}$. We choose the plus branch $v_0^u = H^{-1}$. The normalization condition then allows to solve for the radial components in terms of the other components. The geodesic equation can then be solved perturbatively in the large radius expansion by solving $v_{2}^u$, $v_3^u$, \dots and $v_2^A$, $v_3^A$, \dots while $v_1^u(u,x^A,\lambda)$ and $v_1^A(u,x^A,\lambda)$ are left arbitrary. We introduce for convenience a parameter $\lambda$ that continuously labels distinct geodesics and we denote $s^\mu = \frac{d}{d\lambda} x^\mu$ a generic geodesic deviation vector.

The geodesic deviation equation $\nabla_v \nabla_v s^\mu =R^\mu_{\;\; \alpha\beta\gamma}v^\alpha v^\beta s^\gamma$ then reads explicitly as 
\begin{eqnarray}
\nabla_v \nabla_v s^u &=& \frac{H^2}{r^2} s_1^u+O(r^{-3}),\\
\nabla_v \nabla_v s^r &=& \frac{H^3}{r}(s_1^A q_{AB} v_1^B + H^2 v_1^u s_1^u)+O(r^{-2}),\\
\nabla_v \nabla_v s^A &=& \frac{H^2}{r^2} s_1^A+O(r^{-3}).
\end{eqnarray}

In order to simplify the left-hand side of the geodesic deviation equation, it is useful to express $v^\mu$ in 
Starobinsky coordinates $(T,R,X^{A})$ which are defined from the gauge fixing conditions $g_{TT}=-1$, $g_{TR}=g_{TA}=0$. The time $T$ is the proper time of an asymptotic observer. The change of coordinates between Bondi and Starobinsky coordinates is given at leading order near $\scri^+$, equivalently in the large $r$ or large $T$ limit as 
\begin{eqnarray}
T &=& \frac{\log (H r)}{H}+u +O(r^{-1}), \\
R &=& \frac{e^{-u H}}{H}+O(r^{-1}), \\
X^{A} &=& x^A + O(r^{-2}). 
\end{eqnarray}
At leading order near $\scri^+$,   $v^\mu \partial_\mu = \frac{1}{Hr}\partial_u +Hr \partial_r+\text{subleading}=(1+O(e^{-H T}))\partial_T+O(e^{-H T})\partial_R$. Therefore geodesics can be parameterized by $T$ and we can replace $\nabla_v \nabla_v$ by $\frac{d^2}{dT^2}$. The direction $\partial_R$ is asymptotically orthogonal to the tangent vector of the geodesics $\partial_T$ because $g_{TR}=0$. 

\begin{figure}[tbh]
    \centering
        \includegraphics[width=0.55\textwidth]{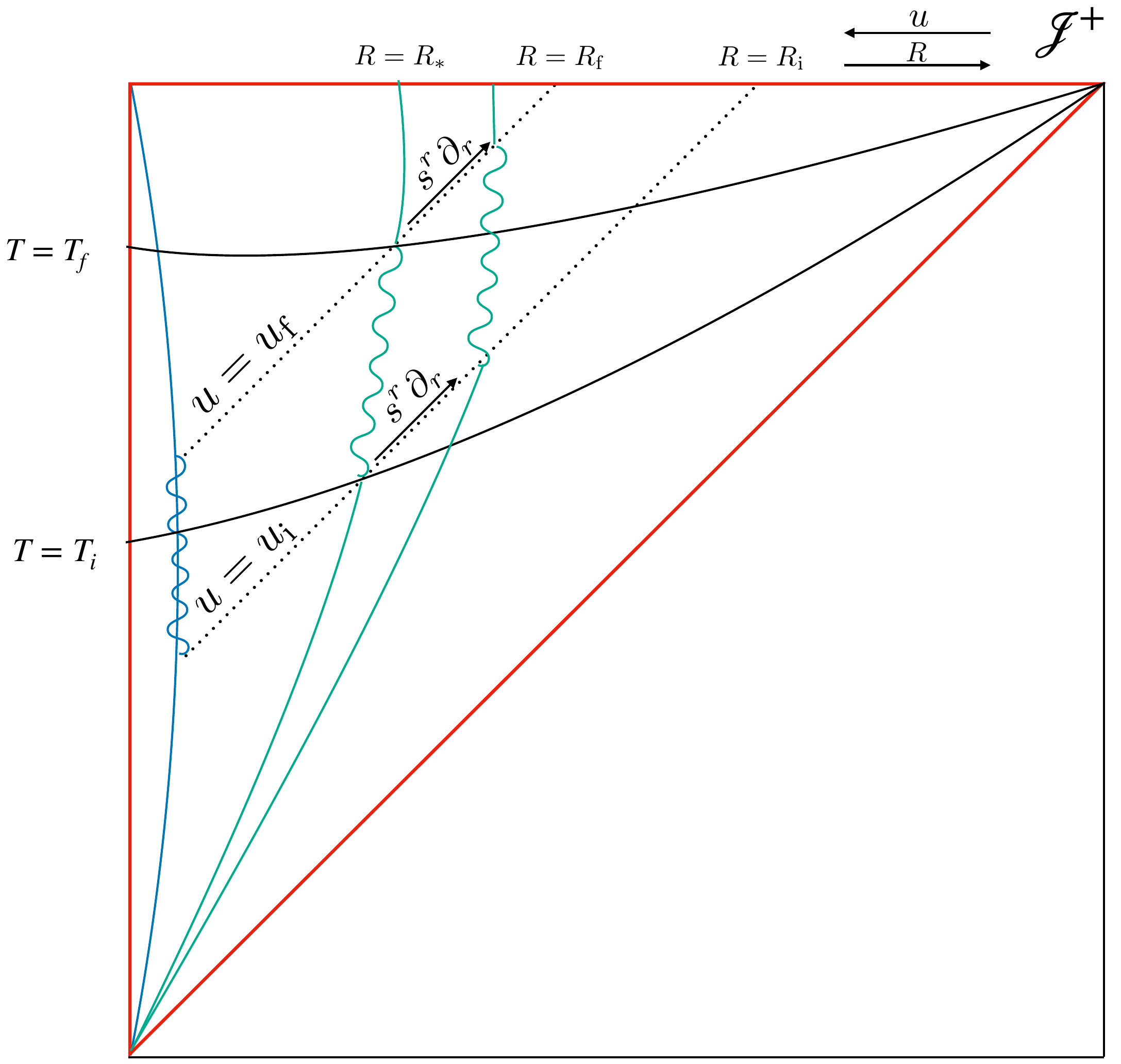}
    \qquad\qquad 
    \caption{In Starobinsky coordinates $(T,R,X^A)$, the geodesic deviation equation can be integrated along time $T$. The displacement memory can be derived from the geodesic deviation component $s^r(T_\text{f},R_*,x^A)-s^r(T_\text{i},R_*,x^A)$ along the Bondi null direction $\partial_r$ or, equivalently in Bondi coordinates, $s^r(u_\text{f},x^A)-s^r(u_\text{i},x^A)$.}\label{Figure5}
\end{figure}

Let us choose a family of geodesics with $v_1^u=0$ and $v_1^A = \bar v_1^A$ at a large cosmological time $r = R e^{HT}\gg H^{-1}$. We take $s_1^u=0$. At leading order close to $\scri^+$ we have 
\bea 
\frac{d^2 s^r}{dT^2} = \frac{H^3}{ R e^{H T}}s_1^A q_{AB}\bar v_1^B ,
\eea
where $q_{AB}(u)=q_{AB}(u(R))$. Upon integration over the direction $T$ (at fixed $R$) and neglecting the integration constants we find
\begin{eqnarray}
    s^r &=& \frac{H}{R e^{H T}} s_1^A q_{AB}\bar v_1^B=\frac{H}{r } s_1^A q_{AB}\bar v_1^B .
\end{eqnarray}

We assume again that gravitational waves reach 
$\scri^+$ only in a finite interval 
$[u_\text{i},u_\text{f}]$ or, equivalently, in 
$[R_\text{f},R_\text{i}]$ with 
$R_\text{f}=H^{-1}e^{-u_\text{f} H} < 
R_\text{i}=H^{-1}e^{-u_\text{i} H}$, see Figure 
\ref{Figure5}. From measuring 
the deviation $s^r( u_\text{f},x^A)$ and 
$s^r(u_\text{i},x^A)$ at a fixed time $r$, one can 
deduce the difference $q_{AB}(u_\text{f},x^A)-q_{AB}
(u_\text{i},x^A)$. We  
can equivalently measure the deviations 
$s^r(T_\text{f},R_*,X^A)$ and 
$s^r(T_\text{i},R_*,X^A)$ at times $T_\text{i}$ 
before the radiation has passed and $T_\text{f}$ 
after the radiation has passed for two geodesics 
separated by an angular deviation $s^A$ at any 
fixed $R_* < R_{\text{f}}$.

\section{Conclusion}
\label{sec:ccl}

In this work we derived the linearized scalar, vector and tensor perturbations in de Sitter background at quadrupolar order in terms of sources localized below the Hubble scale. Contrary to most of the literature on the topic, we kept track consistently all quadrupolar moments, including the odd parity current quadrupole. The derivation of the scalar and vector perturbations are also new with respect to the literature. In the tensorial sector, we  improved upon previous derivations by performing the tail retarded integral explicitly, which allowed to express the tail as a difference between instantaneous terms at retarded time and terms evaluated at the past cosmological horizon. One of our main new result with respect to the existing literature is the consistent quadrupolar truncation, and the check that our generalized harmonic metric reduces to the standard quadrupolar metric \cite{Thorne:1980ru} in the flat limit. In addition, we revised thoroughly the residual gauge transformations in generalized harmonic gauge. 

In a second step, we performed the coordinate transformation of the metric to Bondi gauge and obtained the quadrupolar metric in closed form. Remarkably, all logarithmic terms in the radial expansion cancel and the metric only contains a finite number of terms in the radial expansion, as in the flat limit \cite{Blanchet:2020ngx}, which is correctly recovered. Given the map between Bondi fields and Starobinski/Fefferman-Graham fields \cite{Compere:2019bua}, we confirm the observation that both the boundary metric and the holographic stress-tensor vary upon considering gravitational wave propagation in de Sitter, which invalidates boundary conditions with a fixed boundary metric for generic initial data. Instead, and without loss of generality, we supplemented Bondi gauge with the boundary gauge fixing conditions defined in \cite{Compere:2019bua}, dubbed $\Lambda-$BMS gauge, leading to the $\Lambda$-BMS groupoid as the groupoid of asymptotic symmetries. In the absence of odd parity  multipoles, at locations of $\scri^+$ that are not hit by radiation, the asymptotic symmetric groupoid reduces to an ordinary group, the $\Lambda$-BMS group, which is the equivalent in de Sitter spacetime of the generalized BMS group \cite{Campiglia:2015yka}. Our derivation of the boundary metric in terms of quadrupolar moments implies that any realization of the $\text{dS}_4/\text{CFT}_3$ correspondence  \cite{Strominger:2001pn}, which imposes Dirichlet boundary conditions at $\scri^+$ (consistently with the existence of global conformal symmetries), also severely constraints the bulk initial data. In the absence of incoming radiation from the cosmological horizon, future Dirichlet boundary conditions prevent the existence of any dynamical quadrupolar moment and therefore, presumably, any dynamics.

We finally investigated the imprint of gravitational radiation at leading order at $\scri^+$. We defined the cosmological linear displacement memory effect as a particular geodesic deviation effect due to a change of boundary metric between different spatial locations of $\scri^+$. We proved that this effect is gauge invariant. Furthermore, we identified the precise 10 elements of the $\Lambda$-BMS group that allow to cancel the effect of the quadrupolar energy density and pressure on the boundary metric in one non-radiative region of $\scri^+$. No element of the $\Lambda$-BMS group exists that allows to cancel the effects of the current quadrupole in one non-radiative region. We therefore established that at quadrupolar order the even parity part, but not the odd parity part, of the cosmological linear displacement memory effect originates from a $\Lambda$-BMS transition between vacua related by a  $\Lambda$-BMS group element\footnote{Such a correspondence between memory effects and asymptotic symmetries cannot be achieved for boundary conditions that restrict the asymptotic symmetry group to a finite group such as \cite{Kaminski:2022tum,Bonga:2023eml}.}. We emphasize that this memory effect is a leading order effect in de Sitter based on the flux-balance law of the boundary metric, which becomes trivial in the flat limit, while the flat displacement memory effect is based on the flux-balance law of the (comparatively subleading) Bondi mass aspect.

\section*{Acknowledgements}

G.C. thanks Guillaume Faye for sharing his knowledge and experience of the xAct package and for sharing an unpublished code and he thanks Adrien Fiorucci for comments on the draft. JH thanks Ghanashyam Date and Amitabh Virmani for stimulating conversations. We thank A. Virmani and H. Rana for pointing out a sign mistake in a previous version of this manuscript and we also thank our referees for their relevant comments. G.C. is Senior Research Associate of the F.R.S.-FNRS and acknowledges support from the FNRS research credit J.0036.20F and the IISN convention 4.4503.15. The work of JH is supported in part by MSCA Fellowships CZ - UK2 $(\mbox{reg. n. CZ}.02.01.01/00/22\_010/0008115)$
from the Programme Johannes Amos Comenius co-funded by the European Union. JH also acknowledges the support from Czech Science Foundation Grant 22-14791S. ESK has been
supported respectively by: TUBITAK-2219 Fellowship, TUBITAK Grant No.122R070, TUBITAK Grant No.123R051 throughout the project.

\bibliography{dS_memory}

\providecommand{\href}[2]{#2}\begingroup\raggedright\begin{thebibliography}{10}

\bibitem{PhysRevD.23.347}
A.~H. Guth, {\it Inflationary universe: A possible solution to the horizon and
  flatness problems},  Phys. Rev. D {\bf 23} (Jan, 1981) 347--356.

\bibitem{1998AJ....116.1009R}
A.~G. {Riess}, A.~V. {Filippenko}, P.~{Challis}, A.~{Clocchiatti},
  A.~{Diercks}, P.~M. {Garnavich}, R.~L. {Gilliland}, C.~J. {Hogan}, S.~{Jha},
  R.~P. {Kirshner}, B.~{Leibundgut}, M.~M. {Phillips}, D.~{Reiss}, B.~P.
  {Schmidt}, R.~A. {Schommer}, R.~C. {Smith}, J.~{Spyromilio}, C.~{Stubbs},
  N.~B. {Suntzeff} and J.~{Tonry}, {\it {Observational Evidence from Supernovae
  for an Accelerating Universe and a Cosmological Constant}},  The Astronomical
  Journal {\bf 116} (Sept., 1998) 1009--1038
  [\href{http://arXiv.org/abs/astro-ph/9805201}{{\tt astro-ph/9805201}}].

\bibitem{1999ApJ...517..565P}
S.~{Perlmutter}, G.~{Aldering}, G.~{Goldhaber}, R.~A. {Knop}, P.~{Nugent},
  P.~G. {Castro}, S.~{Deustua}, S.~{Fabbro}, A.~{Goobar}, D.~E. {Groom}, I.~M.
  {Hook}, A.~G. {Kim}, M.~Y. {Kim}, J.~C. {Lee}, N.~J. {Nunes}, R.~{Pain},
  C.~R. {Pennypacker}, R.~{Quimby}, C.~{Lidman}, R.~S. {Ellis}, M.~{Irwin},
  R.~G. {McMahon}, P.~{Ruiz-Lapuente}, N.~{Walton}, B.~{Schaefer}, B.~J.
  {Boyle}, A.~V. {Filippenko}, T.~{Matheson}, A.~S. {Fruchter}, N.~{Panagia},
  H.~J.~M. {Newberg}, W.~J. {Couch} and T.~S.~C. {Project}, {\it {Measurements
  of {\ensuremath{\Omega}} and {\ensuremath{\Lambda}} from 42 High-Redshift
  Supernovae}},  The Astrophysical Journal {\bf 517} (June, 1999) 565--586
  [\href{http://arXiv.org/abs/astro-ph/9812133}{{\tt astro-ph/9812133}}].

\bibitem{PhysRevLett.118.151105}
T.~Regimbau, M.~Evans, N.~Christensen, E.~Katsavounidis, B.~Sathyaprakash and
  S.~Vitale, {\it Digging deeper: Observing primordial gravitational waves
  below the binary-black-hole-produced stochastic background},  Phys. Rev.
  Lett. {\bf 118} (Apr, 2017) 151105.

\bibitem{2023ApJ...951L...8A}
{Nanograv Collaboration, G. Agazie et al.}, {\it {The NANOGrav 15 yr Data Set:
  Evidence for a Gravitational-wave Background}},  The Astrophysical Journal
  Letters {\bf 951} (July, 2023) L8
  [\href{http://arXiv.org/abs/2306.16213}{{\tt 2306.16213}}].

\bibitem{theligoscientificcollaboration2021gwtc3}
T.~L.~S. Collaboration, the Virgo~Collaboration, the KAGRA~Collaboration and
  R.~A. \emph{et al.}, {\it Gwtc-3: Compact binary coalescences observed by
  ligo and virgo during the second part of the third observing run},  2021.

\bibitem{Sathyaprakash:2019nnu}
B.~S. Sathyaprakash {\em et.~al.}, {\it {Cosmology and the Early Universe}},
  \href{http://arXiv.org/abs/1903.09260}{{\tt 1903.09260}}.

\bibitem{vanSon:2021zpk}
L.~A.~C. van Son, S.~E. de~Mink, T.~Callister, S.~Justham, M.~Renzo, T.~Wagg,
  F.~S. Broekgaarden, F.~Kummer, R.~Pakmor and I.~Mandel, {\it {The Redshift
  Evolution of the Binary Black Hole Merger Rate: A Weighty Matter}},
  Astrophys. J. {\bf 931} (2022), no.~1 17
  [\href{http://arXiv.org/abs/2110.01634}{{\tt 2110.01634}}].

\bibitem{Branchesi:2023mws}
M.~Branchesi {\em et.~al.}, {\it {Science with the Einstein Telescope: a
  comparison of different designs}},
  \href{http://arXiv.org/abs/2303.15923}{{\tt 2303.15923}}.

\bibitem{Ford:1977dj}
L.~H. Ford and L.~Parker, {\it {Quantized Gravitational Wave Perturbations in
  Robertson-Walker Universes}},  Phys. Rev. D {\bf 16} (1977) 1601--1608.

\bibitem{Waylen:1978dx}
P.~C. Waylen, {\it {Gravitational Waves in an Expanding Universe}},  Proc. Roy.
  Soc. Lond. A {\bf 362} (1978) 245--250.

\bibitem{Braginsky:1985vlg}
V.~B. Braginsky and L.~P. Grishchuk, {\it {Kinematic Resonance and Memory
  Effect in Free Mass Gravitational Antennas}},  Sov. Phys. JETP {\bf 62}
  (1985) 427--430.

\bibitem{Caldwell:1993xw}
R.~R. Caldwell, {\it {Green's functions for gravitational waves in FRW
  space-times}},  Phys. Rev. D {\bf 48} (1993) 4688--4692
  [\href{http://arXiv.org/abs/gr-qc/9309025}{{\tt gr-qc/9309025}}].

\bibitem{Iliopoulos:1998wq}
J.~Iliopoulos, T.~N. Tomaras, N.~C. Tsamis and R.~P. Woodard, {\it
  {Perturbative quantum gravity and Newton's law on a flat Robertson-Walker
  background}},  Nucl. Phys. B {\bf 534} (1998) 419--446
  [\href{http://arXiv.org/abs/gr-qc/9801028}{{\tt gr-qc/9801028}}].

\bibitem{deVega:1998ia}
H.~J. de~Vega, J.~Ramirez and N.~G. Sanchez, {\it {Generation of gravitational
  waves by generic sources in de Sitter space-time}},  Phys. Rev. D {\bf 60}
  (1999) 044007 [\href{http://arXiv.org/abs/astro-ph/9812465}{{\tt
  astro-ph/9812465}}].

\bibitem{Akhmedov:2010ah}
E.~T. Akhmedov, A.~Roura and A.~Sadofyev, {\it {Classical radiation by
  free-falling charges in de Sitter spacetime}},  Phys. Rev. D {\bf 82} (2010)
  044035 [\href{http://arXiv.org/abs/1006.3274}{{\tt 1006.3274}}].

\bibitem{Chu:2016qxp}
Y.-Z. Chu, {\it {Gravitational Wave Memory In dS$_{4+2n}$ and 4D Cosmology}},
  Class. Quant. Grav. {\bf 34} (2017), no.~3 035009
  [\href{http://arXiv.org/abs/1603.00151}{{\tt 1603.00151}}].

\bibitem{Tolish:2016ggo}
A.~Tolish and R.~M. Wald, {\it {Cosmological memory effect}},  Phys. Rev. D
  {\bf 94} (2016), no.~4 044009 [\href{http://arXiv.org/abs/1606.04894}{{\tt
  1606.04894}}].

\bibitem{Jokela:2022rhk}
N.~Jokela, K.~Kajantie and M.~Sarkkinen, {\it {Gravitational wave memory and
  its tail in cosmology}},  Phys. Rev. D {\bf 106} (2022), no.~6 064022
  [\href{http://arXiv.org/abs/2204.06981}{{\tt 2204.06981}}].

\bibitem{1986RSPTA.320..379B}
L.~{Blanchet} and T.~{Damour}, {\it {Radiative gravitational fields in general
  relativity. I - General structure of the field outside the source}},
  Philosophical Transactions of the Royal Society of London Series A {\bf 320}
  (Dec., 1986) 379--430.

\bibitem{Blanchet:1986dk}
L.~Blanchet, {\it {Radiative gravitational fields in general relativity. 2.
  Asymptotic behaviour at future null infinity}},  Proc. Roy. Soc. Lond. A {\bf
  409} (1987) 383--399.

\bibitem{Ashtekar:2014zfa}
A.~Ashtekar, B.~Bonga and A.~Kesavan, {\it {Asymptotics with a positive
  cosmological constant: I. Basic framework}},  Class. Quant. Grav. {\bf 32}
  (2015), no.~2 025004 [\href{http://arXiv.org/abs/1409.3816}{{\tt
  1409.3816}}].

\bibitem{Ashtekar:2015lxa}
A.~Ashtekar, B.~Bonga and A.~Kesavan, {\it {Asymptotics with a positive
  cosmological constant: III. The quadrupole formula}},  Phys. Rev. D {\bf 92}
  (2015), no.~10 104032 [\href{http://arXiv.org/abs/1510.05593}{{\tt
  1510.05593}}].

\bibitem{Date:2016uzr}
G.~Date and S.~J. Hoque, {\it {Cosmological Horizon and the Quadrupole Formula
  in de Sitter Background}},  Phys. Rev. D {\bf 96} (2017), no.~4 044026
  [\href{http://arXiv.org/abs/1612.09511}{{\tt 1612.09511}}].

\bibitem{Dobkowski-Rylko:2022dva}
D.~Dobkowski-Ry\l{}ko and J.~Lewandowski, {\it {Generalization of the
  quadrupole formula for the energy of gravitational radiation in de Sitter
  spacetime}},  Phys. Rev. D {\bf 106} (2022), no.~4 044047
  [\href{http://arXiv.org/abs/2205.09050}{{\tt 2205.09050}}].

\bibitem{Chu:2015yua}
Y.-Z. Chu, {\it {Transverse traceless gravitational waves in a spatially flat
  FLRW universe: Causal structure from dimensional reduction}},  Phys. Rev. D
  {\bf 92} (2015), no.~12 124038 [\href{http://arXiv.org/abs/1504.06337}{{\tt
  1504.06337}}].

\bibitem{Blanchet:2020ngx}
L.~Blanchet, G.~Comp\`ere, G.~Faye, R.~Oliveri and A.~Seraj, {\it {Multipole
  expansion of gravitational waves: from harmonic to Bondi coordinates}},  JHEP
  {\bf 02} (2021) 029 [\href{http://arXiv.org/abs/2011.10000}{{\tt
  2011.10000}}].

\bibitem{Chrusciel:2021ttc}
P.~T. Chru\'sciel, S.~J. Hoque, M.~Maliborski and T.~Smo\l{}ka, {\it {On the
  canonical energy of weak gravitational fields with a cosmological constant
  $\varLambda \in \mathbb {R}$}},  Eur. Phys. J. C {\bf 81} (2021), no.~8 696
  [\href{http://arXiv.org/abs/2103.05982}{{\tt 2103.05982}}].

\bibitem{He:2017dzb}
X.~He, J.~Jing and Z.~Cao, {\it {Relationship between Bondi\textendash{}Sachs
  quantities and source of gravitational radiation in asymptotically de Sitter
  spacetime}},  Int. J. Mod. Phys. D {\bf 27} (2017), no.~04 1850046
  [\href{http://arXiv.org/abs/1803.05564}{{\tt 1803.05564}}].

\bibitem{Compere:2019bua}
G.~Comp\`ere, A.~Fiorucci and R.~Ruzziconi, {\it {The $\Lambda$-BMS$_4$ group
  of dS$_4$ and new boundary conditions for AdS$_4$}},  Class. Quant. Grav.
  {\bf 36} (2019), no.~19 195017 [\href{http://arXiv.org/abs/1905.00971}{{\tt
  1905.00971}}]. [Erratum: Class.Quant.Grav. 38, 229501 (2021)].

\bibitem{Compere:2020lrt}
G.~Comp\`ere, A.~Fiorucci and R.~Ruzziconi, {\it {The $\Lambda$-BMS$_4$ charge
  algebra}},  JHEP {\bf 10} (2020) 205
  [\href{http://arXiv.org/abs/2004.10769}{{\tt 2004.10769}}].

\bibitem{Barnich:2010eb}
G.~Barnich and C.~Troessaert, {\it {Aspects of the BMS/CFT correspondence}},
  JHEP {\bf 05} (2010) 062 [\href{http://arXiv.org/abs/1001.1541}{{\tt
  1001.1541}}].

\bibitem{Campiglia:2015yka}
M.~Campiglia and A.~Laddha, {\it {New symmetries for the Gravitational
  S-matrix}},  JHEP {\bf 04} (2015) 076
  [\href{http://arXiv.org/abs/1502.02318}{{\tt 1502.02318}}].

\bibitem{Zeldovich:1974aa}
B.~Zeldovich and A.~G. Polnarev., {\it Radiation of gravitational waves by a
  cluster of superdense stars},  Ya. Sov.Astron.Lett. {\bf 18} (1974) 17.

\bibitem{1977ApJ...216..610T}
M.~{Turner}, {\it {Gravitational radiation from point-masses in unbound orbits:
  Newtonian results.}},  Astrophys. J. {\bf 216} (Sep, 1977) 610--619.

\bibitem{1978ApJ...223.1037E}
R.~{Epstein}, {\it {The generation of gravitational radiation by escaping
  supernova neutrinos.}},  Astrophysical J. {\bf 223} (Aug., 1978) 1037--1045.

\bibitem{Blanchet1990}
L.~Blanchet, {\em {Contribution \`a l'\'etude du rayonnement gravitationnel
  \'emis par un syst\`eme isol\'e}}.
\newblock
  \href{http://www2.iap.fr/users/blanchet/TheseHabilitation1990.pdf}{Habilitation
  Thesis}, Universit\'e Pierre et Marie Curie, Paris VI, 1990.
\newblock See Chapter VI, pages 205-214.

\bibitem{Christodoulou:1991cr}
D.~Christodoulou, {\it {Nonlinear nature of gravitation and gravitational wave
  experiments}},  Phys. Rev. Lett. {\bf 67} (1991) 1486--1489.

\bibitem{Wiseman:1991ss}
A.~G. Wiseman and C.~M. Will, {\it {Christodoulou's nonlinear gravitational
  wave memory: Evaluation in the quadrupole approximation}},  Phys. Rev. D {\bf
  44} (1991), no.~10 R2945--R2949.

\bibitem{Th92}
K.~Thorne, {\it Gravitational-wave bursts with memory: The christodoulou
  effect},  Phys. Rev. D {\bf 45} (1992) 520.

\bibitem{Blanchet:1992br}
L.~Blanchet and T.~Damour, {\it {Hereditary effects in gravitational
  radiation}},  Phys. Rev. {\bf D46} (1992) 4304--4319.

\bibitem{Strominger:2014pwa}
A.~Strominger and A.~Zhiboedov, {\it {Gravitational Memory, BMS
  Supertranslations and Soft Theorems}},  JHEP {\bf 01} (2016) 086
  [\href{http://arXiv.org/abs/1411.5745}{{\tt 1411.5745}}].

\bibitem{Satishchandran:2019pyc}
G.~Satishchandran and R.~M. Wald, {\it {Asymptotic behavior of massless fields
  and the memory effect}},  Phys. Rev. D {\bf 99} (2019), no.~8 084007
  [\href{http://arXiv.org/abs/1901.05942}{{\tt 1901.05942}}].

\bibitem{Date:2015kma}
G.~Date and S.~J. Hoque, {\it {Gravitational waves from compact sources in a de
  Sitter background}},  Phys. Rev. D {\bf 94} (2016), no.~6 064039
  [\href{http://arXiv.org/abs/1510.07856}{{\tt 1510.07856}}].

\bibitem{Hinterbichler:2013dpa}
K.~Hinterbichler, L.~Hui and J.~Khoury, {\it {An Infinite Set of Ward
  Identities for Adiabatic Modes in Cosmology}},  JCAP {\bf 01} (2014) 039
  [\href{http://arXiv.org/abs/1304.5527}{{\tt 1304.5527}}].

\bibitem{Hui:2018cag}
L.~Hui, A.~Joyce and S.~S.~C. Wong, {\it {Inflationary soft theorems revisited:
  A generalized consistency relation}},  JCAP {\bf 02} (2019) 060
  [\href{http://arXiv.org/abs/1811.05951}{{\tt 1811.05951}}].

\bibitem{Mirbabayi:2016xvc}
M.~Mirbabayi and M.~Simonovi\'c, {\it {Weinberg Soft Theorems from Weinberg
  Adiabatic Modes}},  \href{http://arXiv.org/abs/1602.05196}{{\tt 1602.05196}}.

\bibitem{Hamada:2018vrw}
Y.~Hamada and G.~Shiu, {\it {Infinite Set of Soft Theorems in Gauge-Gravity
  Theories as Ward-Takahashi Identities}},  Phys. Rev. Lett. {\bf 120} (2018),
  no.~20 201601 [\href{http://arXiv.org/abs/1801.05528}{{\tt 1801.05528}}].

\bibitem{Compere:2017wrj}
G.~Comp\`ere, R.~Oliveri and A.~Seraj, {\it {Gravitational multipole moments
  from Noether charges}},  JHEP {\bf 05} (2018) 054
  [\href{http://arXiv.org/abs/1711.08806}{{\tt 1711.08806}}].

\bibitem{Assassi:2012zq}
V.~Assassi, D.~Baumann and D.~Green, {\it {On Soft Limits of Inflationary
  Correlation Functions}},  JCAP {\bf 11} (2012) 047
  [\href{http://arXiv.org/abs/1204.4207}{{\tt 1204.4207}}].

\bibitem{Ashtekar:2015ooa}
A.~Ashtekar, B.~Bonga and A.~Kesavan, {\it {Gravitational waves from isolated
  systems: Surprising consequences of a positive cosmological constant}},
  Phys. Rev. Lett. {\bf 116} (2016), no.~5 051101
  [\href{http://arXiv.org/abs/1510.04990}{{\tt 1510.04990}}].

\bibitem{Hoque:2018byx}
S.~J. Hoque and A.~Virmani, {\it {On Propagation of Energy Flux in de Sitter
  Spacetime}},  Gen. Rel. Grav. {\bf 50} (2018), no.~4 40
  [\href{http://arXiv.org/abs/1801.05640}{{\tt 1801.05640}}].

\bibitem{Ashtekar:2017ydh}
A.~Ashtekar and B.~Bonga, {\it {On a basic conceptual confusion in
  gravitational radiation theory}},  Class. Quant. Grav. {\bf 34} (2017),
  no.~20 20LT01 [\href{http://arXiv.org/abs/1707.07729}{{\tt 1707.07729}}].

\bibitem{Thorne:1980ru}
K.~S. Thorne, {\it {Multipole Expansions of Gravitational Radiation}},  Rev.
  Mod. Phys. {\bf 52} (1980) 299--339.

\bibitem{Damour:1990gj}
T.~Damour and B.~R. Iyer, {\it {Multipole analysis for electromagnetism and
  linearized gravity with irreducible cartesian tensors}},  Phys. Rev. D {\bf
  43} (1991) 3259--3272.

\bibitem{Poole:2018koa}
A.~Poole, K.~Skenderis and M.~Taylor, {\it {(A)dS$\mathbf{_4}$ in Bondi
  gauge}},  Class. Quant. Grav. {\bf 36} (2019), no.~9 095005
  [\href{http://arXiv.org/abs/1812.05369}{{\tt 1812.05369}}].

\bibitem{Chrusciel:2020rlz}
P.~T. Chru\'sciel, S.~J. Hoque and T.~Smo\l{}ka, {\it {Energy of weak
  gravitational waves in spacetimes with a positive cosmological constant}},
  Phys. Rev. D {\bf 103} (2021), no.~6 064008
  [\href{http://arXiv.org/abs/2003.09548}{{\tt 2003.09548}}].

\bibitem{Bonga:2023eml}
B.~Bonga, C.~Bunster and A.~P\'erez, {\it {Gravitational radiation with
  $\Lambda>0$}},  \href{http://arXiv.org/abs/2306.08029}{{\tt 2306.08029}}.

\bibitem{Strominger:2001pn}
A.~Strominger, {\it {The dS / CFT correspondence}},  JHEP {\bf 10} (2001) 034
  [\href{http://arXiv.org/abs/hep-th/0106113}{{\tt hep-th/0106113}}].

\bibitem{Kaminski:2022tum}
W.~Kami\'nski, M.~Kolanowski and J.~Lewandowski, {\it {Symmetries of the
  asymptotically de Sitter spacetimes}},  Class. Quant. Grav. {\bf 39} (2022),
  no.~19 195009 [\href{http://arXiv.org/abs/2205.14093}{{\tt 2205.14093}}].

\end{thebibliography}\endgroup
\end{document}